\documentclass[10pt,twocolumn,letterpaper]{article}

\usepackage[pagenumbers]{cvpr} 

\usepackage[dvipsnames]{xcolor}

\newcommand{\mypar}[1]{\vspace{-0.5cm}\paragraph{#1}}
\newcommand{\myparb}[1]{\vspace{0.0cm}\paragraph{#1}}

\newcommand{\bea}{\begin{eqnarray}}
\newcommand{\eea}{\end{eqnarray}}

\definecolor{cvprblue}{rgb}{0.21,0.49,0.74}
\usepackage[pagebackref,breaklinks,colorlinks,citecolor=cvprblue]{hyperref}

\def\paperID{1625} 
\def\confName{CVPR}
\def\confYear{2024}

\title{APISR: Anime Production Inspired Real-World Anime Super-Resolution}

\author{Boyang Wang\textsuperscript{1} \quad
Fengyu Yang\textsuperscript{1,2*} \quad
Xihang Yu\textsuperscript{1} \quad
Chao Zhang\textsuperscript{3} \quad
Hanbin Zhao\textsuperscript{3\dag}
\vspace{5mm} \\
\textsuperscript{1}University of Michigan \quad 
\textsuperscript{2}Yale University \quad 
\textsuperscript{3}Zhejiang University\\
}

\begin{document}
\maketitle
{\let\thefootnote\relax\footnotetext{{{\dag} Corresponding author}}}
{\let\thefootnote\relax\footnotetext{{{*} works done at University of Michigan}}}
\begin{abstract}
While real-world anime super-resolution (SR) has gained increasing attention in the SR community, existing methods still adopt techniques from the photorealistic domain. In this paper, we analyze the anime production workflow and rethink how to use characteristics of it for the sake of the real-world anime SR. First, we argue that video networks and datasets are not necessary for anime SR due to the repetition use of hand-drawing frames. Instead, we propose an anime image collection pipeline by choosing the least compressed and the most informative frames from the video sources. Based on this pipeline, we introduce the Anime Production-oriented Image (API) dataset. In addition, we identify two anime-specific challenges of distorted and faint hand-drawn lines and unwanted color artifacts. We address the first issue by introducing a prediction-oriented compression module in the image degradation model and a pseudo-ground truth preparation with enhanced hand-drawn lines. In addition, we introduce the balanced twin perceptual loss combining both anime and photorealistic high-level features to mitigate unwanted color artifacts and increase visual clarity. We evaluate our method through extensive experiments on the public benchmark, showing our method outperforms state-of-the-art anime dataset-trained approaches. The code is available at \href{https://github.com/Kiteretsu77/APISR}{\textit{https://github.com/Kiteretsu77/APISR}}.

\vspace{-0.3cm}
\end{abstract}
\section{Introduction}
\vspace{-0.1cm}
\label{sec:intro}

As an important subdiscipline of real-world super-resolution (SR), anime SR focuses on restoring and enhancing low-quality low-resolution (LR) anime visual art images and videos to high-quality high-resolution (HR) forms. It has demonstrated significant practical impacts in the fields of entertainment and commerce~\cite{wu2022animesr,tuo2023learning, xu2022transformer,wang2021realesrgan,wang2024vcisr}.
An emerging line of work has addressed the problem by extending SR networks to capture multi-scale information or learning an adaptive degradation model~\cite{wu2022animesr, tuo2023learning}.
We argue these methods lack understanding of the anime domain as their techniques are directly transplanted from the photorealistic SR approach.

\begin{figure}[t]

     \vspace{-0.1cm}
  \centering
  \includegraphics[width= 1.0\columnwidth]{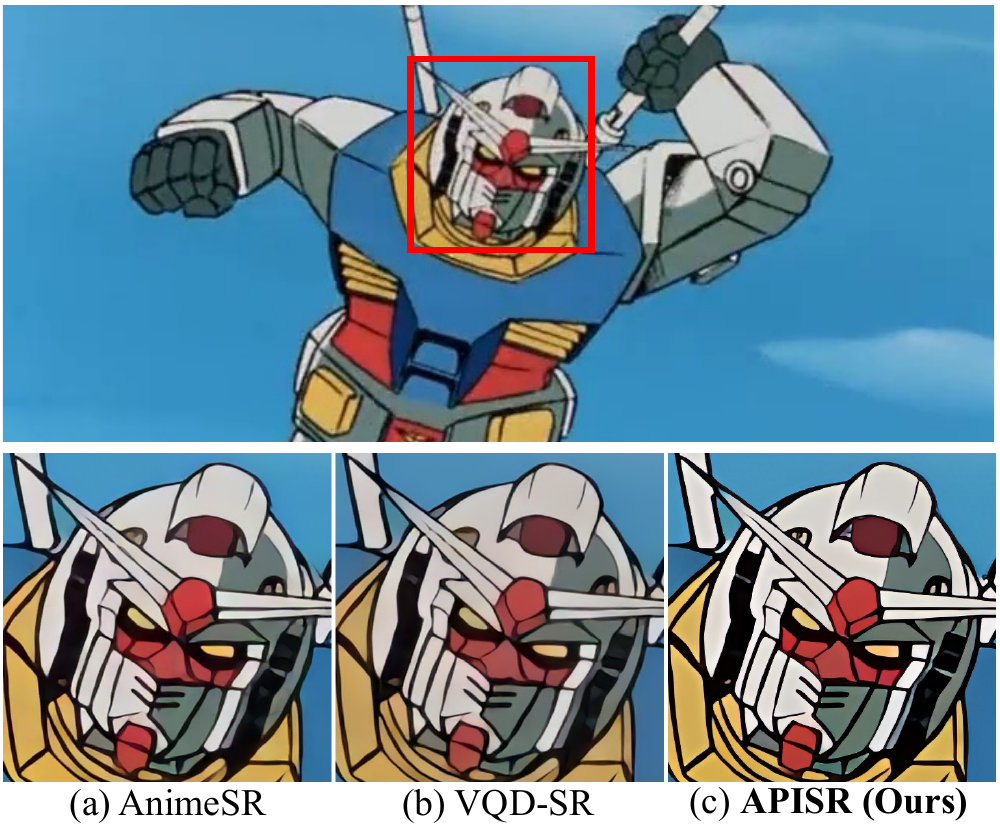}
  
   \vspace{-0.2cm}
   
       \caption{\textbf{Comparisons between proposed APISR and other SOTA anime SR methods.} Ours present clearer and sharper hand-drawn lines, better restoration with more natural details, and do not present unwanted color artifacts. \textbf{Zoom in for best view.}}

   \label{fig:teaser}

   \vspace{-0.5cm}
\end{figure}
In this paper, we thoroughly analyze the anime production process, exploring ways to leverage its unique aspects for practical applications in anime SR.
The production workflow first starts with hand-drawing sketches on paper, which are then colorized and enhanced by computer-generated imagery (CGI) processing~\cite{zhao2022cartoon}.
Then, these processed sketches are concatenated into a video. 
Due to the fact that the drawing process is extremely labor-intensive and human eyes are not sensitive to motions~\cite{shen2022enhanced, chen2022improving}, it is a standard practice to reuse a single image across multiple consecutive frames when forming the video.
This procedure in production motivates us to rethink whether it is necessary and efficient to use video networks and video datasets to train SR networks in the anime domain.

To this end, we explore the use of image-based methods and datasets as a unified super-resolution and restoration framework for both anime images and videos.
Creating an image dataset allows us more flexibility to exclusively choose the least-compressed video frames as our potential dataset pool, rather than gathering sequential frames that contain temporal distortions to create a video dataset.
Furthermore, by forming an image dataset, we can selectively focus on the most informative frames, as anime videos typically possess less information than photorealistic videos.
If we randomly crop a patch from an anime image, there is a high probability that it is a monochromatic area signifying a lack of information.
In light of these phenomena, we introduce an anime image collection pipeline that focuses on keyframes in video, along with an image complexity assessment-based selection criteria. This method is designed to identify and select the least-compressed and the most informative images from video sources.
Using our pipeline, we propose \textbf{A}nime \textbf{P}roduction-oriented \textbf{I}mage (API) dataset for SR training.

\begin{figure}[t]
  \centering
    
    \vspace{-0.0cm}
    
  \includegraphics[width= 1.0\columnwidth]{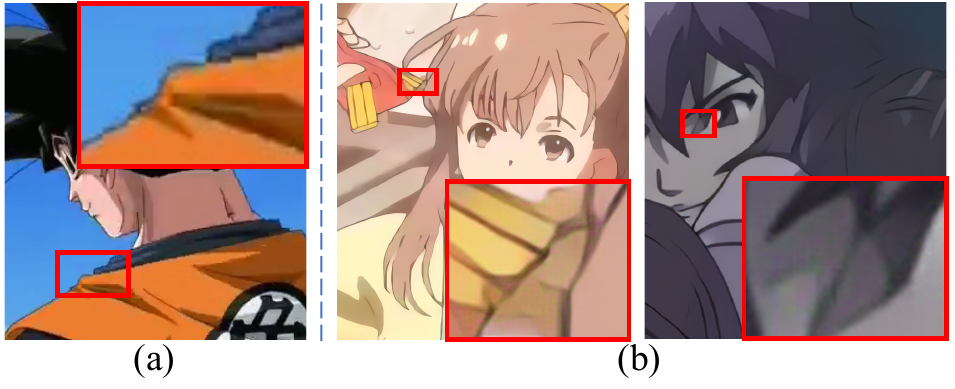}
    
    \vspace{-0.3cm}
    
   \caption{
   \textbf{We identify two new anime-specific challenges}: (a) Distorted and faint hand-drawn lines frequently appear in real-world anime images. (b) Unwanted color artifacts in AnimeSR~\cite{wu2022animesr} and VQD-SR~\cite{tuo2023learning}. \textbf{Zoom in for the best view.}
   }

    \vspace{-0.4cm}
   
   \label{fig:sharp_and_color}
\end{figure}

In addition, we identify two new anime-specific challenges for real-world SR tasks. 
First, in anime production, the clarity of hand-drawn lines is a highly emphasized detail~\cite{wu2022animesr, jiang2023scenimefy, carrillo2023diffusart} as shown in Fig.~\ref{fig:sharp_and_color} a, but hand-drawn lines are easily weakened due to compression in internet transmission and physical aging in production. This deterioration at the edges of lines exerts a substantial negative impact on the visual effects.
To address this, we start from the perspective of restoration and enhancement.
Concretely, we propose a prediction-oriented compression module in the image degradation model to simulate compression in internet transmission such that the model trained with this self-supervised method can restore hand-drawn line distortions. 
In addition, we propose a ground-truth (GT) enhancement approach to enhance faint, aging hand-drawn lines, by merging hand-drawn lines extracted from the overly sharpened GT images.

Second, we realize an issue of unwanted color artifacts in anime images, which is a consequence of employing the GAN-based SR networks~\cite{goodfellow2014generative} (see Fig.~\ref{fig:sharp_and_color} b).
These artifacts are presented as irregularly shaped colored spots with varying intensities that are scattered randomly across generated images, which significantly undermines visual perception. 
We attribute this issue to the reason that image features of perceptual loss are trained on the photorealistic image datasets, which is inconsistent in the anime domain.
To mitigate this issue, we conduct a comprehensive study of perceptual loss and introduce balanced twin perceptual loss, which assembles perceptual features from both the photorealistic domain and the anime domain by a balanced layer scaling distribution.

Thus, we summarize our contributions as follows:
\begin{itemize}
    \item We propose a novel anime dataset curation pipeline that is capable of collecting the least compressed and the most informative anime images from video sources.
    \item We propose an image degradation model to deal with harder compression restoration challenges, especially for hand-drawn line distortions, and the first methodologies in the anime domain to attentively enhance faint hand-drawn lines.
    \item We realize and address the unwanted color artifacts in GAN-based SR network training caused by the domain inconsistency of the perceptual loss. 
    \item  We thoroughly evaluate our method on the real-world anime SR dataset and show that our method outperforms state-of-the-art anime dataset-trained SR approaches by a large margin with only 13.3\% training sample complexity of the prior work. 
\end{itemize}

\section{Realated Work}
\label{sec:related_work}

\myparb{Real-World Super-Resolution.}
Classical SR methods~\cite{dong2015image, chan2021basicvsr, xiao2020space, xiao2021space} typically employs a straightforward approach, using a single bicubic downsampling operation to convert high-resolution (HR) ground-truth (GT) images into their low-resolution (LR) counterparts. Classical image restoration methods~\cite{liang2021swinir, xiao2023dive, xiao2023cutmib, li2023efficient} train different weights for different tasks. 
In contrast, real-world SR is dedicated to implementing a sophisticated degradation model by one model weight to restore the diverse degradations found in the real-world scenario, such as blurring, noise, and compression~\cite{Ji_2020_CVPR_Workshops, wang2021realesrgan, zhang2021designing, liang2021swinir, li2023efficient, wang2024vcisr}.

Generic degradation model design can be broadly classified into two categories: explicit models~\cite{Ji_2020_CVPR_Workshops, wang2021realesrgan, zhang2021designing, li2023efficient, wang2024vcisr} and implicit models~\cite{wu2022animesr,tuo2023learning, yuan2018unsupervised}. Explicit degradation models employ kernels and mathematical formulas to simulate real-world degradation processes. On the other hand, the implicit degradation models focus on training neural networks to capture the distribution of real-world degradations. 
Nevertheless, implicit models face challenges of interpretability and scalability. The efficacy of implicit models lacks a clear rationale, and adapting them to new domains requires the creation of bespoke datasets and extra training complexity.

\mypar{Anime Processing.}
Anime represents a distinctive form of visual art, often characterized by exaggerated visual representation. 
Creators of anime typically start by sketching line art, followed by 2D and 3D animation techniques, which include elements like colorization, CGI effects, and frame interpolation. 
Notably, recent research in the realm of anime has garnered substantial attention, \eg, AI painting with anime content~\cite{zhang2023adding, guo2023animatediff, huang2023anifacedrawing}, vectorization of anime images~\cite{zhang2009vectorizing, yao2016manga}, anime interpolation and inbetweening~\cite{siyao2021deep, chen2022improving, siyao2023deep}, anime sketch colorization~\cite{carrillo2023diffusart, cao2023animediffusion, wang2023coloring, zhang2021user, dai2024learning}, 3D representation~\cite{siyao2022animerun, chen2023panic}, and anime domain adaptation~\cite{jiang2023scenimefy}.

AnimeSR (NeurIPS 2022)~\cite{wu2022animesr} and VQD-SR (ICCV 2023)~\cite{tuo2023learning} are two recent representative studies in the domain of real-world anime super-resolution tasks. 
However, they have not fully addressed the unique challenges of low-level anime restoration. This includes the faint hand-drawn lines and domain inconsistency in the training of GAN-based networks.
This paper conducts a comprehensive exploration of several meticulously crafted approaches to the anime SR domain.

\section{Proposed Method}

\subsection{Anime Production-Oriented Image SR Dataset}
In this section, we present the \textbf{API} (\textbf{A}nime \textbf{P}roduction-oriented \textbf{I}mage) SR dataset and its curation workflow. This curation leverages the characteristics of anime videos to select the least compressed and the most informative frames.

\begin{figure}[t]
  \centering
  \vspace{-0.2cm}
  \includegraphics[width= 1.0\columnwidth]{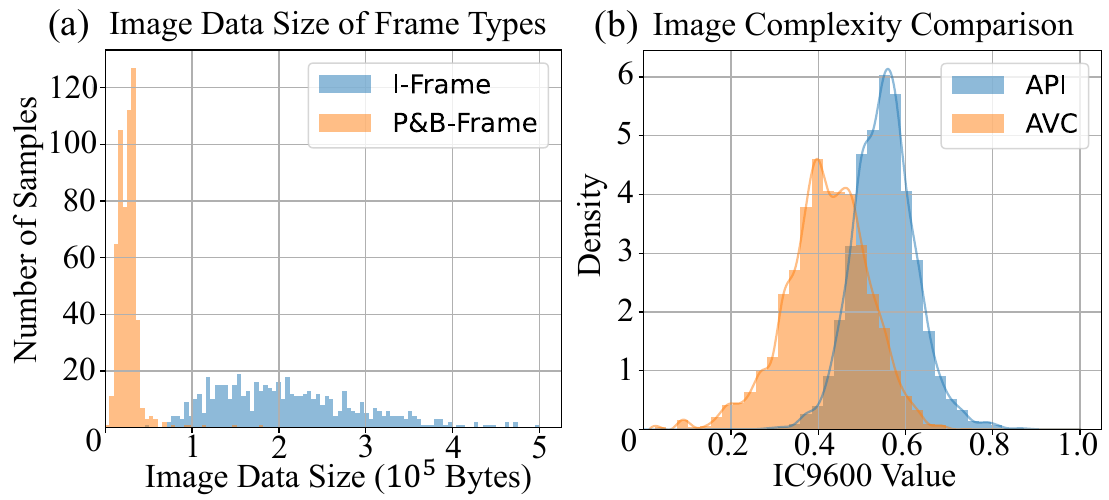}

  \vspace{-0.3cm}
  
   \caption{Histogram of (a) the average image data size comparison between I-Frames and Non-I-Frames (P and B-Frame) in collected video sources and (b) image complexity~\cite{feng2022ic9600} comparison between proposed API and AVC~\cite{wu2022animesr} dataset.}

   \vspace{-0.0cm}
   \label{fig:histogram}
\end{figure}

\begin{figure}[t]
  \centering
  \includegraphics[width= 1.0\columnwidth]{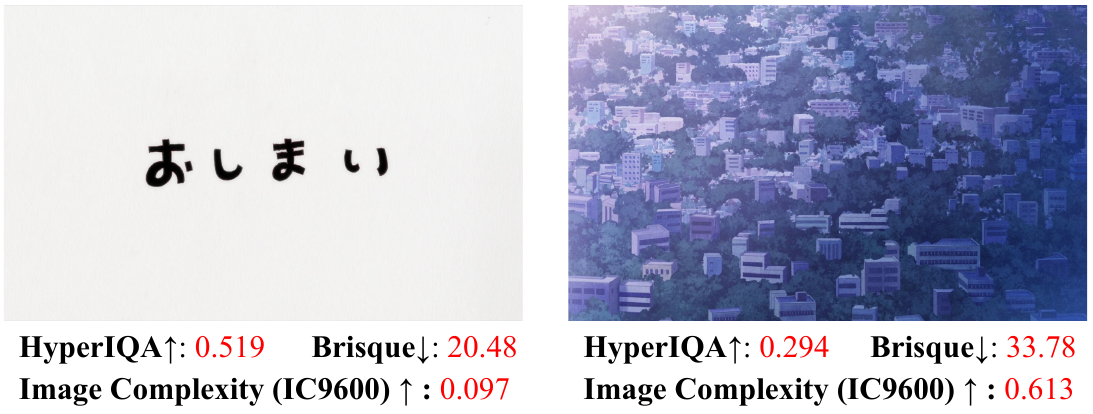}

  \vspace{-0.2cm}
  
   \caption{
      Image Quality Assessment (IQA) with HyperIQA~\cite{su2020blindly} and Brisque~\cite{mittal2012no} vs. Image Complexity Assessment (ICA) with IC9600~\cite{feng2022ic9600}. IQA favors simple scenes and gives low scores to images with strong CGI. However, ICA is the opposite.
   }

   \vspace{-0.4cm}
   
    \label{fig:IQA_IC}
\end{figure}
\begin{figure}[t]

  \centering
  \vspace{-0.1cm}
  \includegraphics[width= 1.0\columnwidth]{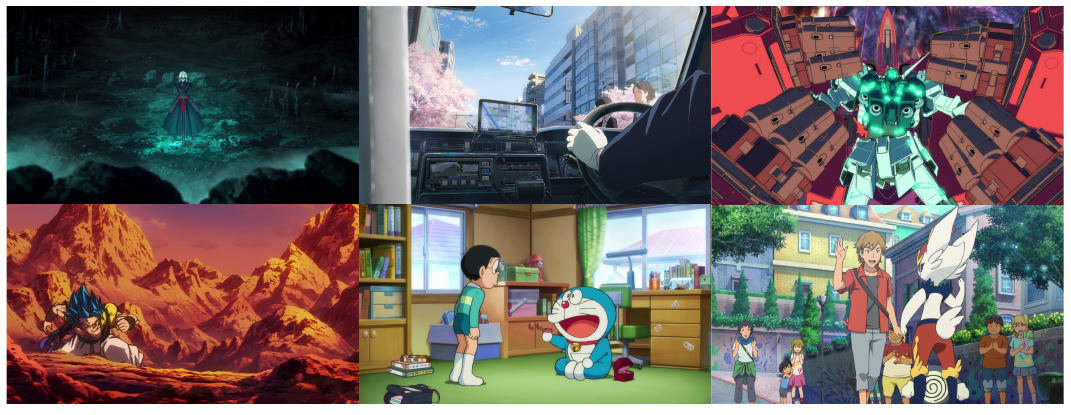}
  \vspace{-0.6cm}
  
   \caption{
   \textbf{Samples of API Super-Resolution Dataset.} API includes versatile CGI effects scenes (\eg, different lightning and special effects) and presents high image complexity. }
   
   \vspace{-0.5cm}
   \label{fig:api_dataset}
\end{figure}
\begin{figure*}[t]
  \centering
  
  \vspace{-0.4cm}
  
  \includegraphics[width=\linewidth]{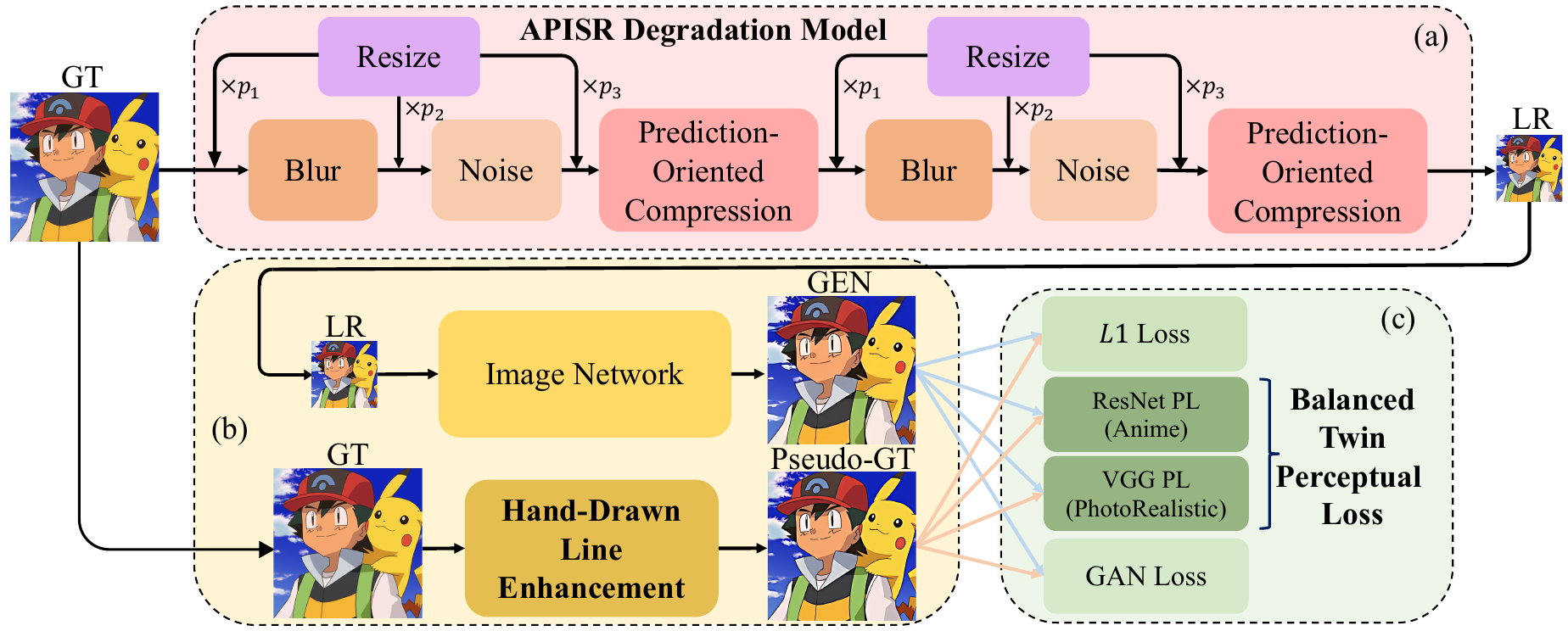}
  
  \vspace{-0.3cm}
  
   \caption{
   \textbf{The overview of our proposed methods.} 
   (a) We proposed a prediction-oriented compression module in the degradation model to simulate versatile compression degradations for a single image input (detailed in Sec.~\ref{sec:degradation}). Proposed shuffled resize module is randomly positioned to augment the representation of the degradation model. 
   (b) GT images are augmented with proposed hand-drawn line enhancement to promote the generation of images with sharpened line edge details in training (detailed in Sec.~\ref{sec:enhance}). 
   (c) Proposed balanced twin perceptual loss avoids unwanted color artifacts in GAN network training (detailed in Sec.~\ref{sec:perceptual}).
   }
   
   \vspace{-0.4cm}
   \label{fig:pipeline_main}
\end{figure*}

\noindent\textbf{I-Frame-based Image Collection.}
AnimeSR introduces AVC-Train, the first video-based anime SR dataset, but they overlook the impact of compression during the collection process, which leads VQD-SR to propose a post-processing technique to enhance the dataset. 
Instead, we propose a novel method to select the least compressed frames from the source level with minimum effort.

All videos on the internet are compressed and encapsulated with a video compression standard (\eg, H.264~\cite{schwarz2007overview} and H.265~\cite{sullivan2012overview}) for a trade-off between the quality and the data size.
There are numerous video compression standards, each with a complex engineering system, but they share a similar backbone design. 
This characteristic motivates us to find the pattern that the compression quality assigned to each frame is different. 
Video compression designates some keyframes, known as I-Frames, as individual units for compression. Empirically, I-Frames are the first frame of scene-changing scenarios. These I-Frames are allocated with a high data size budget. 
On the contrary, a higher compression ratio requires non-I-Frames, namely P-Frames and B-Frames, to take I-Frames as the reference during compression, which introduces temporal distortions. As shown in Fig.~\ref{fig:histogram} a, among the anime videos we collect, I-Frames on average have a much higher data size than other non-I-Frames, which genuinely stand for higher quality. 
Thus, we use \textit{ffmpeg}, a video processing tool, to extract all I-Frames from the video source as an initial pool.

\noindent\textbf{Image Complexity-based Selection.}
To further select idealistic images from the I-Frames pool, we need some criteria.
A straightforward method involves following AVC-Train to use the Image Quality Assessment (IQA) to rank and choose frames with better scores. 
However, IQA ranking does not prefer anime images with CGI effects but favors simple scenes with little information (see Fig.~\ref{fig:IQA_IC}).
Thus, we argue that image complexity assessment (ICA) is a better option in the anime domain.

ICA evaluates the level of intricacy in an image by scoring the amount and variety of details present. Compared to IQA, ICA demonstrates greater robustness against changes in saturation, lightning, contrast, and motion blurring.
The ICA metric we use is a recent rising analysis network, IC9600~\cite{feng2022ic9600}.
In the anime domain, employing ICA presents two primary advantages.
First, many scenes in anime videos are characteristically monotonous (as exemplified in Fig.~\ref{fig:IQA_IC} left), where the majority of pixels lack significant information in training. IQA favors these simple images and gives higher score compared to other images, but ICA enables the exclusion of these scenes, which, in turn, contributes to a reduced training sample complexity.
Second, ICA is more adept at identifying meaningful scenes within anime production, especially those featuring CGI effects, such as the dark scene in Fig.~\ref{fig:IQA_IC} right. These are scenarios where IQA methods typically falter. By collecting versatile scenes, the network training can become more robust in handling complex real-world anime inputs.

\noindent\textbf{API Dataset.}
We began by manually sourcing 562 high-quality anime videos. From these, we extracted all I-Frames as an initial selection pool. 
Utilizing the image complexity assessment method mentioned above, we then selected the top 10 highest-scoring frames from the I-Frames pool of each video.
After discarding inappropriate images (\eg, nudity, violence, abnormality, and anime images mixed with photorealistic content), 3,740 high-quality images are obtained as our proposed dataset. 
Example images are shown in Fig.~\ref{fig:api_dataset}. 
Moreover, as shown in Fig.\ref{fig:histogram} b, the density of high image complexity scored frames of our API dataset is remarkably superior to that of AVC-Train. More analysis and data can be found in the supplementary materials.

\noindent\textbf{720P Back-to-Original Production Resolution.}
While studying the anime production pipeline, we observed that most anime productions follow a 720P format (with an image height of 720 pixels). However, in real-world scenarios, anime is often falsely upscaled to 1080P or other formats, for the sake of standardizing multimedia formats. 
We empirically find that rescaling all anime images back to the original 720P can provide feature density envisioned by the creators with more compact anime hand-drawn lines and CGI information.

\subsection{An Anime Practical Degradation model}
\label{sec:degradation}

In the real-world SR, the design of the degradation model is of great importance.
Based on the high-order degradation model~\cite{wang2021realesrgan} and a recent image-based video compression restoration model~\cite{wang2024vcisr}, we propose two improvements to restore distorted hand-drawn lines and versatile compression artifacts and to augment the representation of the degradation model. Our degradation model is shown in Fig.~\ref{fig:pipeline_main} a.

\begin{figure}[t]
  \centering
  \includegraphics[width= 1.0\columnwidth]{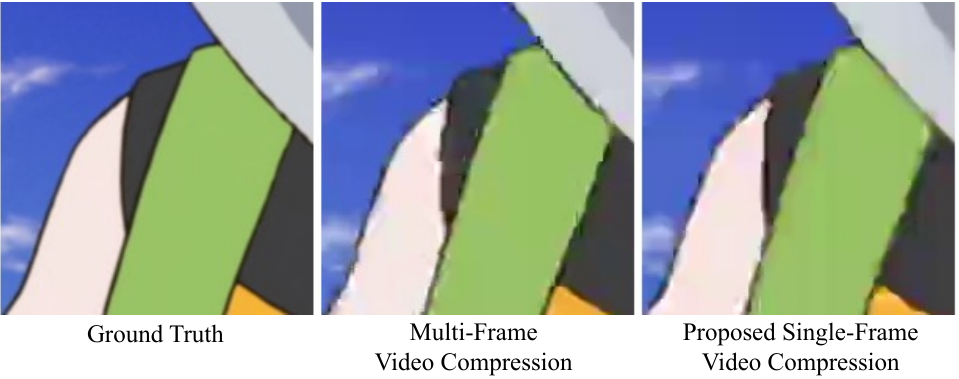}

  \vspace{-0.3cm}
  
   \caption{
        H.264~\cite{schwarz2007overview} compression of regular multi-frame video compression and our proposed single-frame compression. They exert similar degradations (\eg, distortion to hand-drawn lines).
   }

   \vspace{-0.4cm}
   \label{fig:compression}
\end{figure}

\noindent\textbf{Prediction-Oriented Compression.}
Utilizing the image degradation model presents a challenge in the anime restoration of video compression artifacts.
This is because previous real-world image SR methods employ JPEG, an old but widely-used image compression standard, as the sole compression module in the image degradation model. 
JPEG performs repetitive and independent compression on all encoding units, without considering the existence of other units. However, video compression algorithms, for higher compression ratios, apply prediction algorithms to search for a reference with similar pixel content and only compress their differences (residual), thereby reducing information entropy.
Prediction algorithms can search their reference spatially (intra-prediction) or temporally (inter-prediction). Regardless of the category, the intrinsic cause of distortion comes from the misalignment in residual due to prediction limitation.

Hence, we argue that artifacts equivalent to real-world video compression artifacts can be synthesized using a single image input in conjunction with a prediction-oriented compression algorithm (\eg, WebP~\cite{si2016research} and H.264). 
The need for genuinely sequential frames is not necessary.
To this end, we design a prediction-oriented compression module within the image degradation model. This module requires video compression algorithms to compress inputs on a single-frame basis.
Compared to VCISR~\cite{wang2024vcisr}, we don't need multiple frames for one turn of execution of compression.
This methodology is theoretically reasonable and practically viable from an engineering perspective. With a single-frame input, video compression trivially applies intra-prediction to compress the frame without using its inter-prediction functionality.
Utilizing this approach, the image degradation model is capable of synthesizing compression artifacts akin to those observed in conventional multi-frame video compression as shown in Fig.~\ref{fig:compression}. 
Subsequently, by feeding these synthesized images into the image SR network, the system can effectively learn the patterns of versatile compression artifacts and engage in the restoration.

\noindent\textbf{Shuffled Resize Module.}
Degradation models in the real-world SR domain consider blurring, resize, noise, and compression modules. Blurring, noise, and compression are real-world artifacts that can be synthesized with clear mathematical models or algorithms. 
However, the logic of the resize module is entirely different. 
Resize is not a part of natural image generation but is introduced solely for SR-paired dataset purposes. 
Given this notion, we believe that previous fixed resize module is not very suitable. We propose a more robust and effective solution, which involves randomly placing resize operations at various orders in the degradation model.

\subsection{Anime Hand-Drawn Lines Enhancement}
\label{sec:enhance}

To enhance faint hand-drawn lines, directly employing global methods, such as modifying the degradation model or sharpening the entire GT, is not an ideal approach, as the network cannot learn with attention to hand-drawn line changes.
Thus, we choose to extract sharpened hand-drawn line information and merge it back with GT to form pseudo-GT. 
By introducing this attentively enhanced pseudo-GT to SR training, the network can generate sharpened hand-drawn lines without the need to introduce additional neural network modules or separate post-processing networks.

To extract hand-drawn lines, a direct approach is to apply a sketch extraction model. However, current learning-based sketch extraction is often characterized by a style transfer to the reference image, which distorts hand-drawn line details and encompasses unrelated pixel content (\eg, shadows and edges of CGI effects).  
Consequently, we need a more granular, pixel-by-pixel methodology to extract hand-drawn lines.
Thus, we utilize XDoG~\cite{winnemoller2012xdog}, a pixel-by-pixel Gaussian kernel-based sketch extraction algorithm, to extract edge maps from the sharpened GT. 
Nevertheless, XDoG edge maps are marred by excessive noise, containing outlier pixels and fragmented line representations.
To address this ill-posed issue, we propose an outlier filtering technique coupled with a custom-designed passive dilation method (detailed in the supplementary materials). In this way, we yield a more coherent and undisturbed representation of hand-drawn lines.

\begin{figure}[t]

    \vspace{-0.1cm}
    
  \centering
  \includegraphics[width= 1.0\columnwidth]{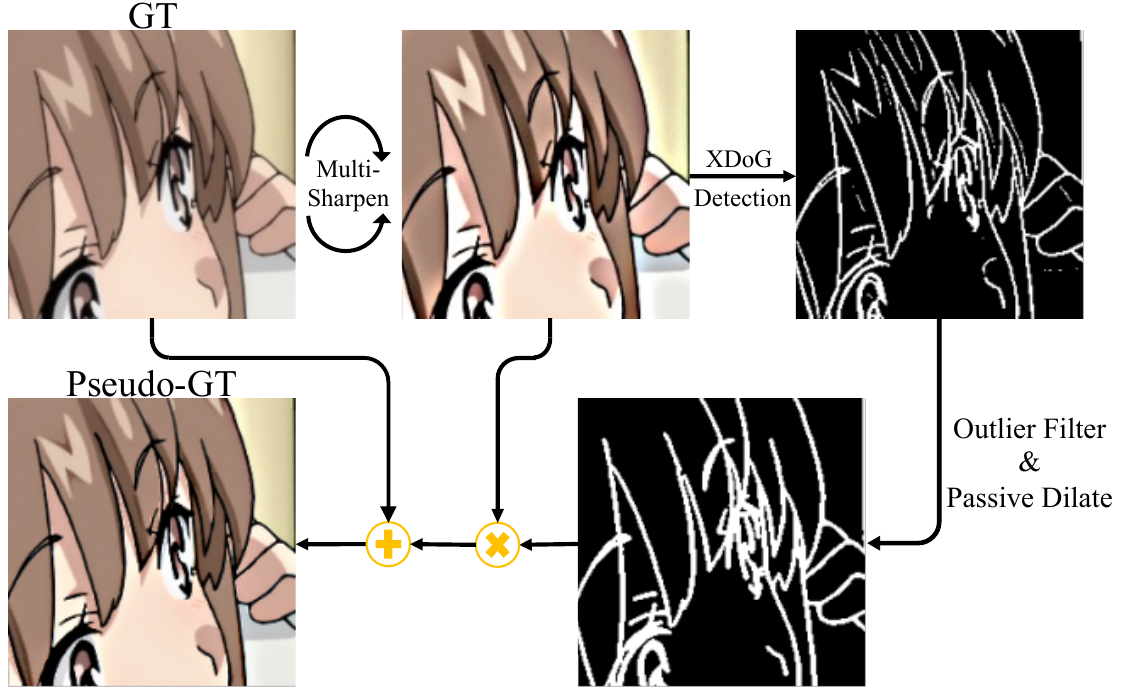}
  \vspace{-0.5cm}
  
   \caption{\textbf{Anime Hand-Drawn Lines Enhancement Pipeline.}}
   
   \vspace{-0.4cm}
   \label{fig:anime_sharpen}
\end{figure}

We empirically find that overly sharpened pre-processed GT makes the hand-drawn line margins more noticeable than other unrelated shadow edge details, which makes the outlier filter easier to distinguish their differences.
Thus, we propose three rounds of unsharp masking to the GT first.
To sum up, the formula is as follows:
\begin{gather}
    I_{\text{Sharp}} = f^n(I_{\text{GT}}), \\
    I_{\text{Map}} = h(g(I_{\text{Sharp}})), \\
    I_{\text{pseudo-GT}} = I_{\text{Sharp}} \cdot I_{\text{Map}} + I_{\text{GT}} \cdot (1 - I_{\text{Map}}),
\end{gather}
where $f$ is the sharpening function that recursively executes $n$ times, $g$ denotes XDoG edge detection and $h$ stands for post-processing techniques of passive dilation with outlier filtering. $I_{\text{Map}}$ is a binary value map.
The visual pipeline is shown in Fig.~\ref{fig:anime_sharpen}.

\subsection{Balanced Twin Perceptual Loss for Anime}
\label{sec:perceptual}

The existence of unwanted color artifacts is attributed to the inconsistent dataset domain in training between the generator and perceptual loss. 
Currently, most SR models trained with GAN, including AnimeSR and VQD-SR, use the same ImageNet~\cite{deng2009imagenet} pre-trained VGG~\cite{simonyan2014very} network as the perceptual loss.
However, anime content, particularly those mixed with CGI and extensive illustrations, differs significantly from photorealistic features in ImageNet. 
To tackle this problem, we investigate perceptual loss and the subsequent improvements made in their following works.

\begin{figure}[t]
  \centering
  \vspace{-0.1cm}
  
  \includegraphics[width= 1.0\columnwidth]{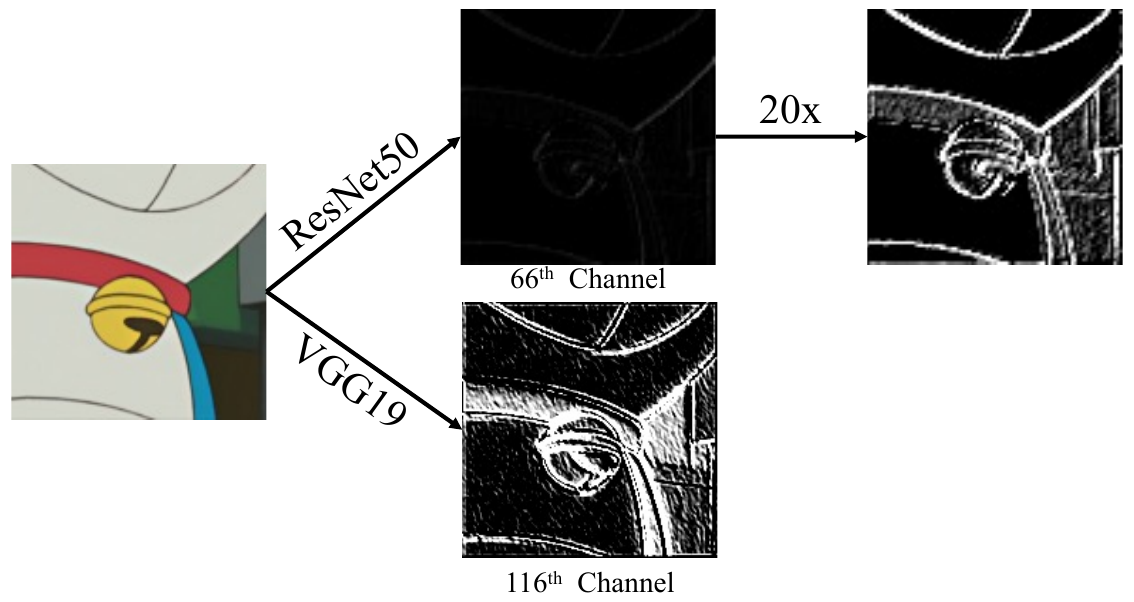}
  
  \vspace{-0.3cm}
  
   \caption{
   The second middle-layer feature outputs comparison between VGG19 used by photo-realistic perceptual loss~\cite{johnson2016perceptual} and ResNet50 used by anime recognition task~\cite{danbooru2018resnet, branwen2019danbooru2019}. With scaling, ResNet50 presents a similar intensity as the VGG outputs.
   }

   \vspace{-0.4cm}

   \label{fig:layer}
\end{figure}

The core idea behind perceptual loss is to utilize high-level features (\eg, segmentation, classification, recognition) to complement low-level pixel features by comparing middle-layer feature outputs.
In this regard, we employ a pre-trained ResNet50~\cite{he2016identity, danbooru2018resnet} on anime object classification task with Danbooru~\cite{branwen2019danbooru2019} dataset, a substantial and rich tagging anime illustration database.
Since the pre-trained network is ResNet50 instead of VGG, we propose a similar middle-layer comparison (detailed in the supplementary material).
Overall, the formula is as follows:
\bea
L^{\phi}_{ResNet}(\hat{y},y) = {\sum_j}\frac{w_j}{C_j H_j W_j}\left| \phi_j(\hat{y}) - \phi_j(y) \right|,
\eea
where $y$ and $\hat{y}$ are the pseudo-GT by Sec.~\ref{sec:enhance} and the generated images. $\phi_j$ represents the perceptual function that returns $j$th layer output of ResNet50. $C_j$, $H_j$, and $W_j$ are dimensions of the layer output and $w_j$ is the scaling factor for each layer. 
There are 5 middle-layer feature outputs, which is the same quantity as VGG-based perceptual loss.
We also observe that the intensity of shallow feature layers in ResNet50 is very weak (see Fig.~\ref{fig:layer}). To resemble a similar intensity balance as the VGG, we apply a high $w_j$ to the early layers, which leads to stable training.

Notably, introducing the ResNet-based perceptual loss as the sole perceptual loss can solve unwanted color artifacts and lead to quantitative improvements. 
However, there may be instances of poor visual results. 
This is attributed to the inherent bias in the Danbooru dataset, where most images are character faces or relatively simple illustrations.
Hence, we seek a tradeoff by using real-world features as an auxiliary primer to guide the ResNet-based perceptual loss in training. This approach results in visually appealing images and also resolves the unwanted color issue. The overall loss function for our GAN training is defined as follows:
\bea
L = \alpha L_1 + \beta L_{per} + \gamma L_{adv}, \\
L_{per} = L_{ResNet} + \delta L_{VGG},
\eea
where $L_1$, $L_{VGG}$, and $L_{adv}$ are L1 pixel loss, photorealistic VGG-based perceptual loss, and the adversarial loss. $\alpha$, $\beta$, $\gamma$ and $\delta$ are weight parameters.

\section{Experiment}
\vspace{-0.1cm}
\subsection{Implementation Details}
In our experiment, we employ our proposed API dataset as the training dataset for the image network. The image network we utilize is a tiny version of GRL~\cite{li2023efficient} with the nearest convolution upsample module (detailed in the supplementary). 

To train the GAN, we follow the same two-stage training approach as prior works~\cite{wang2021realesrgan, wang2018esrgan, wu2022animesr, zhang2021designing,chan2022investigating}. In the first stage, we train the network with L1 pixel loss for 300K iterations. In the second stage, we introduce our balanced twin perceptual loss and the adversarial loss, conducting an additional 300K iterations. The weights of $\{\alpha, \beta, \gamma, \delta \}$ are $\{1, 0.5, 0.2, 1\}$ respectively. 
The layer weight of perceptual loss is $\{0.1, 20, 25, 1, 1\}$ for ResNet and $\{0.1, 1, 1, 1, 1\}$ for VGG.
Our discriminator is the same three-scale patch discriminator~\cite{isola2017image, wang2018high, miyato2018spectral} as in AnimeSR~\cite{wu2022animesr} and VQD-SR~\cite{tuo2023learning}.
We use the Adam optimizer~\cite{kingma2014adam} with a learning rate of $2\times 10^{-4}$ in the first stage and $1\times 10^{-4}$ in the second stage. A learning rate decay is applied every 100K iterations in both stages. The entire training process was carried out on one Nvidia RTX 4090, with HR patch sizes set at 256x256 and a batch size of 32.

As for the degradation model, we perform degradation on the whole HR image first rather than directly on a cropped patch as in previous works~\cite{wang2021realesrgan, li2023efficient, zhang2021designing, wang2024vcisr}. Within the degradation model, noise and blurring are configured identically to Real-ESRGAN~\cite{wang2021realesrgan}, and the first prediction-oriented compression is implemented with JPEG~\cite{wallace1992jpeg} and WebP~\cite{si2016research}. The second prediction-oriented compression includes AVIF~\cite{han2021technical}, JPEG~\cite{wallace1992jpeg}, WebP~\cite{si2016research}, and single-frame compression of MPEG2~\cite{mitchell1996mpeg}, MPEG4~\cite{avaro2000mpeg}, H.264~\cite{schwarz2007overview}, and H.265~\cite{sullivan2012overview}. 
The probability of placing the resize module is equally divided among all positions.
Specific parameter settings can be found in our supplementary materials.

\begin{figure*}[t]
\vspace{-0.1cm}
  \centering
  
  \includegraphics[width=\linewidth]{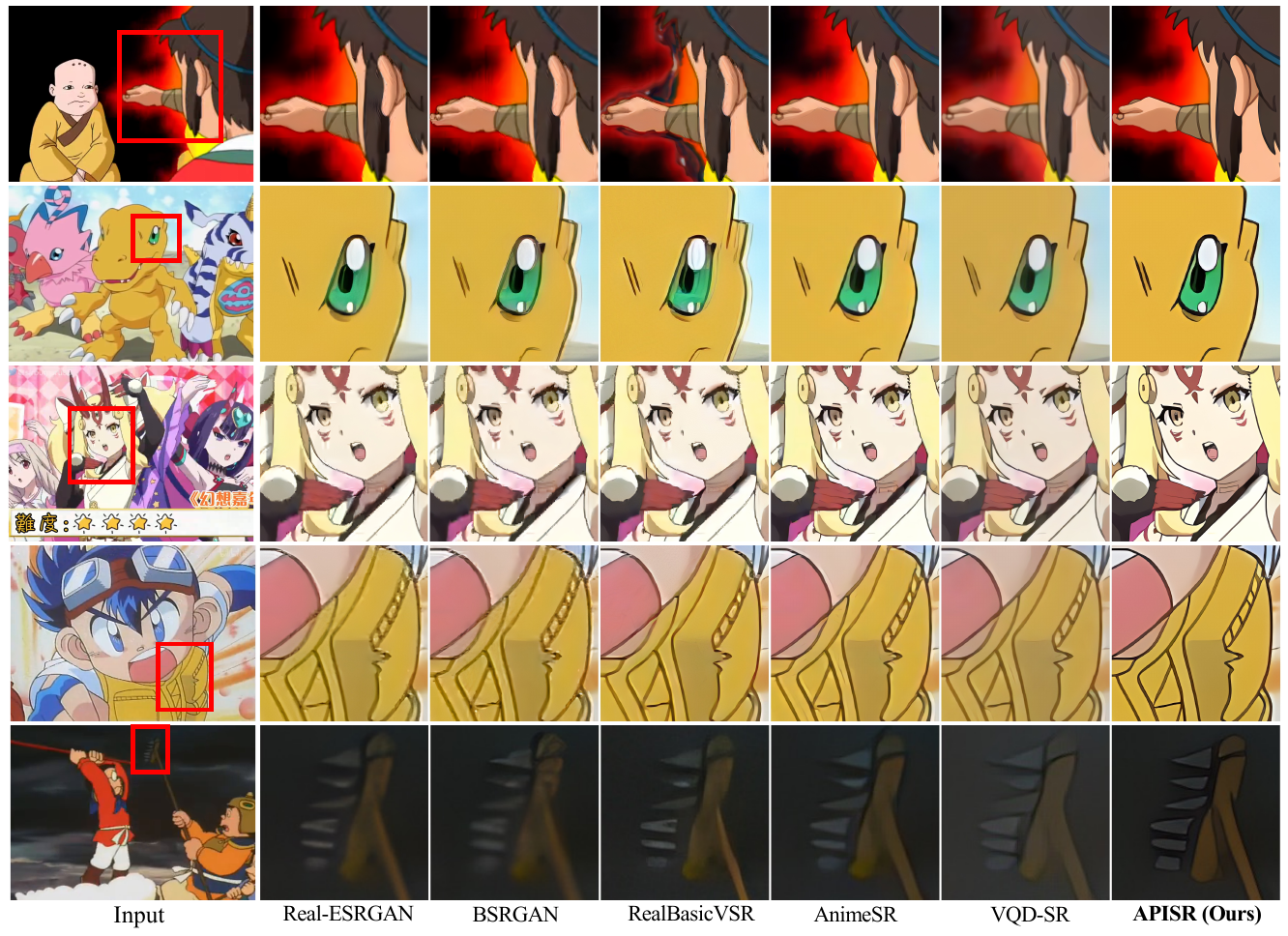}

  \vspace{-0.4cm}
  
   \caption{
   Qualitative comparisons on AVC-RealLQ~\cite{wu2022animesr} for $4\times$ scaling. \textbf{Zoom in for the best view.}
   }
    \label{fig:visualization_main}
   \vspace{-0.4cm}
\end{figure*}

\subsection{Comparisons with State-of-the-art Methods}
We compare our APISR quantitatively and qualitatively with other SOTA real-world image and video SR methods, which include Real-ESRGAN~\cite{wang2021realesrgan}, BSRGAN~\cite{zhang2021designing}, 
Real-BasicVSR~\cite{chan2022investigating}, AnimeSR~\cite{wu2022animesr}, and VQD-SR~\cite{tuo2023learning}.

\begin{table}[t]
	\footnotesize
	\centering
	\caption{\small{Quantitative comparisons on AVC-RealLQ~\cite{wu2022animesr}. \textbf{Bold} text indicates the best performance. (`$*$' denotes fine-tune on animation videos from ~\cite{wu2022animesr})}}
	\label{tab:quantitative_comparison}
	\vspace{-0.2cm}
 \scalebox{0.9}{
		\begin{tabular}{l||c|c|c|c}
			\hline
			\textbf{Methods} & Params $\downarrow$ & NIQE$\downarrow$ & MANIQA$\uparrow$ &CLIPIQA$\uparrow$ \\
			\hline\hline
			Real-ESRGAN*~\cite{wang2021realesrgan} & 16.70 & 8.281 & 0.381 & - \\
			BSRGAN*~\cite{zhang2021designing} & 16.70 & 8.632 & 0.376 & - \\ 
			RealBasicVSR*~\cite{chan2022investigating}  & 6.30 & 8.621 & 0.362 & -  \\
			AnimeSR~\cite{wu2022animesr} & 1.50 & 8.109 & 0.462 & 0.539\\
            VQD-SR~\cite{tuo2023learning} & 1.47 & 8.202 & 0.464 & 0.567 \\
                \textbf{APISR (Ours)} & \textbf{1.03} & \textbf{6.719} & \textbf{0.514} & \textbf{0.711} \\
			\hline
	   \end{tabular}
        }
 \vspace{-0.5cm}
\end{table}

\mypar{Quantitative Comparison.} 
Following previous real-world SR works~\cite{wang2021realesrgan, chan2022investigating, wu2022animesr, tuo2023learning, ji2020real}, we conduct inference on low-quality LR datasets to generate high-quality HR images and evaluate them using no-reference metrics. The scaling factor is 4 for all methods. 
To validate the effectiveness of our approach, our evaluation is based on AVC-RealLQ~\cite{wu2022animesr}, which has 46 video clips each with 100 frames. 
This dataset is the only known dataset designed for real-world anime SR testing. 
For no-reference metrics, we employ the same metrics used in VQD-SR and AnimeSR, which are NIQE~\cite{mittal2012making} and MANIQA~\cite{yang2022maniqa}.
We also incorporate other SOTA learning-based image quality assessment metrics like CLIPIQA~\cite{wang2023exploring}. All metrics are based on pyiqa\cite{pyiqa} library.

As shown in Tab.~\ref{tab:quantitative_comparison}, our model has the smallest network size, 1.03M parameters, but has SOTA performance in all metrics among all image and video-based methods. 
Apart from the various proposed methods that contribute to our success, special acknowledgment is due to the design of the prediction-oriented compression model, which enables us to train image datasets and image networks to restore video compression degradations. 
Meanwhile, it is worth mentioning that we achieved the result with only 13.3\% and 25\% of the training sample complexity of AnimeSR~\cite{wu2022animesr} and VQD-SR~\cite{tuo2023learning}. 
This is especially thanks to the introduction of image complexity assessment in dataset curation which selects informative images to increase the efficacy of learning the representation of anime images. 
Further, we require zero training on the degradation model due to the explicit degradation model we design.

\mypar{Qualitative Comparison.} 
As shown in Fig.~\ref{fig:visualization_main}, APISR greatly improves the visual quality than other methods. In restoring heavily compressed images, our model exhibits exceptional proficiency than all other methods, as exemplified in the first row, where we have much fewer ringing artifacts. 
Moreover, owing to the proposed hand-drawn lines enhancement, our generated images manifest increased line density and clarity as observed in the second row. 
In addressing various twisted lines and shadow artifacts, our model outperforms others in effective restoration, evidenced by the third and fourth rows. This is thanks to our improvement to the image degradation model where we provide a robust restoration capability on compression and resize functionality.
Meanwhile, due to our proposed balanced twin perceptual loss, images generated by our GAN network do not show unwanted color artifacts as in AnimeSR and VQD-SR, which can be seen in the fifth row. Further, thanks to the versatile scenes collected in our proposed dataset, we are capable of achieving effective restoration in dark scenes. More visual results can be found in the supplementary materials.

\begin{table}[]
\footnotesize
\centering
\caption{Ablation study results of different training datasets. IQA stands for image quality assessment. ICA stands for image complexity assessment.}
\vspace{-0.2cm}
 \label{tab:ablation_dataset}
\setlength{\tabcolsep}{1mm}{
\begin{tabular}{@{}lcccc@{}}
\toprule
\textbf{Dataset}                 & \scriptsize NIQE$\downarrow$ & \scriptsize MANIQA$\uparrow$ & \scriptsize CLIPIQA$\uparrow$ \\ \midrule
AVC-Train~\cite{wu2022animesr}      & 7.681     & 0.476     & 0.658        \\ 
Random Select                & 8.006     & 0.446     & 0.625        \\ 
I-Frame + IQA Select         & 7.876     & 0.493     & 0.675        \\
I-Frame + ICA Select          & 6.912     & 0.499     & 0.683         \\
I-Frame + ICA Select + 720P Rescale          & \textbf{6.719}     & \textbf{0.514}     & \textbf{0.711}         \\\bottomrule
\end{tabular}
}
\vspace{-0.1cm}
\end{table}

\begin{table}[]
\footnotesize
\centering
\caption{Ablation study results of different degradation model.}
\vspace{-0.2cm}
 \label{tab:ablation_degradation}
\setlength{\tabcolsep}{1mm}{
\begin{tabular}{@{}lccc@{}}
\toprule
\textbf{Degradation Model}                       & \scriptsize NIQE$\downarrow$ & \scriptsize MANIQA$\uparrow$ &\scriptsize CLIPIQA$\uparrow$\\ \midrule
High-Order~\cite{wang2021realesrgan}  & \textbf{6.667}     & 0.483     & 0.663         \\
Random Order~\cite{zhang2021designing}  & 6.975     & 0.491     & 0.674        \\
Prediction-Oriented Compression         & 7.133     & 0.506     & 0.709         \\
Compression + Shuffled Resize           & 6.719     & \textbf{0.514}     & \textbf{0.711}         \\ \bottomrule
\end{tabular}}
\vspace{-0.1cm}
\end{table}

\begin{table}[]
\footnotesize
\centering
\caption{
Ablation study results of hand-drawn lines enhancement denoted as \textbf{Sharpen} and twin perceptual loss denoted as \textbf{APL}.
}
\vspace{-0.2cm}
 \label{tab:ablation_sharpen_perceptual}
\setlength{\tabcolsep}{1mm}{
\begin{tabular}{@{}lccc@{}}
\toprule
                                & \scriptsize NIQE$\downarrow$ & \scriptsize MANIQA$\uparrow$ &\scriptsize CLIPIQA$\uparrow$       \\ \midrule
Plain                        & 7.351     & 0.501     & 0.689         \\
Plain + Sharpen    & 7.182     & 0.504     & 0.707         \\
Plain + Sharpen + APL     & 6.835     & 0.512     & 0.708          \\
Plain + Sharpen + APL + Balanced Scale     & \textbf{6.719}     & \textbf{0.514}     & \textbf{0.711}         \\\bottomrule
\end{tabular}}

\vspace{-0.4cm}

\end{table}

\subsection{Ablation Study}
In this section, we conduct ablation studies to evaluate the substantial impact of our proposed dataset, degradation model, and hand-drawn lines enhancement with balanced twin perceptual loss.
The inference dataset is still AVC-RealLQ~\cite{wu2022animesr}.
Visual comparisons are presented in the supplementary materials.

\mypar{Impact of the Dataset.} 
As shown in Tab.~\ref{tab:ablation_dataset}, we substitute our API training dataset with several alternatives for comparative analysis: AVC-Train~\cite{wu2022animesr}, frames randomly selected from the same video source as our API, a collection of I-Frames with IQA selection, and a collection of I-Frames with ICA selection. 
For a fair comparison, we keep a similar intensity of the training dataset size.
If we take the AVC-Train video training dataset as an image dataset to train, we include temporal distorted images and less informative frames, which makes the performance hard to compete with the model trained with API in all metrics.
Randomly selected image datasets perform poorly because they lack attention to high-quality frames in videos. 
With our I-Frame collection, we take off temporally distorted frames and choose the least compressed frames, but IQA-based selection limits the performance. 
With the same training iterations and conditions, the dataset selected by ICA-based criteria leads to an improvement over the dataset by IQA-based selection.
With the 720P rescaling method, anime images have more compact hand-drawn lines and CGI information than falsely upscaled versions, and this back-to-original thinking boosts the performance in all metrics.

\mypar{Degradation Model.}
As shown in Tab.~\ref{tab:ablation_degradation}, to validate the superiority of our degradation model, we replace our proposed degradation model with the high-order degradation model from the Real-ESRGAN~\cite{wang2021realesrgan} and random order degradation model from BSRGAN~\cite{zhang2021designing}, which share certain similarity as our methods. Our degradation model with prediction-oriented compression model reaches an outstanding improvement in MANIQA~\cite{yang2022maniqa} and CLIPIQA~\cite{wang2023exploring} metrics. 
With our shuffled resize design, our network becomes more robust to versatile real-world SR scenarios and the performance can move one step further, especially the NIQE~\cite{mittal2012making} metrics.

\mypar{Benefits of proposed Enhancement and Perceptual Loss.}
As shown in Tab.~\ref{tab:ablation_sharpen_perceptual}, we compare our model with the plain version that is not trained with proposed hand-drawn lines enhancement and balanced twin perceptual loss. The introduction of our hand-drawn lines enhancement presents a significant improvement on CLIPIQA~\cite{wang2023exploring}. 
When we append ResNet perceptual loss in GAN training, it shows outstanding improvement in NIQE~\cite{mittal2012making}. 
Further, with the proposed scaling on the early layers of the ResNet perceptual loss part, two perceptual losses have reached a stable balance and the performance moves one step further.
This proves that a perceptual loss that is compatible with the anime domain is very insightful and instructive.

\section{Conclusion}

In this paper, we thoroughly utilize the characteristics of anime production knowledge and fully leverage it to enrich and enhance anime SR. 
To be specific, we propose a high-quality and informative anime production-oriented image (API) SR dataset with a novel dataset curation design. 
To restore and enhance hand-drawn lines, we propose an image degradation model to restore video compression artifacts and a pseudo-GT enhancement strategy. We further address unwanted color artifacts by introducing a network trained with high-level anime tasks to construct a balanced twin perceptual loss. Extensive experiment results demonstrate our superiority over existing SOTA methods, where we can restore harder real-world low-quality anime images.

{
    \small
    \bibliographystyle{ieeenat_fullname}
    \bibliography{main}

\begin{thebibliography}{69}
\providecommand{\natexlab}[1]{#1}
\providecommand{\url}[1]{\texttt{#1}}
\expandafter\ifx\csname urlstyle\endcsname\relax
  \providecommand{\doi}[1]{doi: #1}\else
  \providecommand{\doi}{doi: \begingroup \urlstyle{rm}\Url}\fi

\bibitem[Agustsson and Timofte(2017)]{agustsson2017ntire}
Eirikur Agustsson and Radu Timofte.
\newblock Ntire 2017 challenge on single image super-resolution: Dataset and study.
\newblock In \emph{Proceedings of the IEEE conference on computer vision and pattern recognition workshops}, pages 126--135, 2017.

\bibitem[Avaro et~al.(2000)Avaro, Eleftheriadis, Herpel, Rajan, and Ward]{avaro2000mpeg}
Olivier Avaro, Alexandros Eleftheriadis, Carsten Herpel, Ganesh Rajan, and Liam Ward.
\newblock Mpeg-4 systems: overview.
\newblock \emph{Signal Processing: Image Communication}, 15\penalty0 (4-5):\penalty0 281--298, 2000.

\bibitem[Baas(2019)]{danbooru2018resnet}
Matthew Baas.
\newblock Danbooru2018 pretrained resnet models for pytorch.
\newblock \url{https://rf5.github.io}, 2019.
\newblock Accessed: DATE.

\bibitem[Branwen and Gokaslan(2019)]{branwen2019danbooru2019}
Gwern Branwen and Aaron Gokaslan.
\newblock Danbooru2019: A large-scale crowdsourced and tagged anime illustration dataset.
\newblock \emph{Danbooru2017}, 2019.

\bibitem[Cao et~al.(2023)Cao, Meng, Mok, Liu, Lee, and Li]{cao2023animediffusion}
Yu Cao, Xiangqiao Meng, PY Mok, Xueting Liu, Tong-Yee Lee, and Ping Li.
\newblock Animediffusion: Anime face line drawing colorization via diffusion models.
\newblock \emph{arXiv preprint arXiv:2303.11137}, 2023.

\bibitem[Carrillo et~al.(2023)Carrillo, Cl{\'e}ment, Bugeau, and Simo-Serra]{carrillo2023diffusart}
Hernan Carrillo, Micha{\"e}l Cl{\'e}ment, Aur{\'e}lie Bugeau, and Edgar Simo-Serra.
\newblock Diffusart: Enhancing line art colorization with conditional diffusion models.
\newblock In \emph{Proceedings of the IEEE/CVF Conference on Computer Vision and Pattern Recognition}, pages 3485--3489, 2023.

\bibitem[Chan et~al.(2021)Chan, Wang, Yu, Dong, and Loy]{chan2021basicvsr}
Kelvin~CK Chan, Xintao Wang, Ke Yu, Chao Dong, and Chen~Change Loy.
\newblock Basicvsr: The search for essential components in video super-resolution and beyond.
\newblock In \emph{Proceedings of the IEEE/CVF conference on computer vision and pattern recognition}, pages 4947--4956, 2021.

\bibitem[Chan et~al.(2022)Chan, Zhou, Xu, and Loy]{chan2022investigating}
Kelvin~CK Chan, Shangchen Zhou, Xiangyu Xu, and Chen~Change Loy.
\newblock Investigating tradeoffs in real-world video super-resolution.
\newblock In \emph{Proceedings of the IEEE/CVF Conference on Computer Vision and Pattern Recognition}, pages 5962--5971, 2022.

\bibitem[Chen and Mo(2022)]{pyiqa}
Chaofeng Chen and Jiadi Mo.
\newblock {IQA-PyTorch}: Pytorch toolbox for image quality assessment.
\newblock [Online]. Available: \url{https://github.com/chaofengc/IQA-PyTorch}, 2022.

\bibitem[Chen and Zwicker(2022)]{chen2022improving}
Shuhong Chen and Matthias Zwicker.
\newblock Improving the perceptual quality of 2d animation interpolation.
\newblock In \emph{European Conference on Computer Vision}, pages 271--287. Springer, 2022.

\bibitem[Chen et~al.(2023)Chen, Zhang, Shi, Wang, Zhu, Song, An, Kristjansson, Yang, and Zwicker]{chen2023panic}
Shuhong Chen, Kevin Zhang, Yichun Shi, Heng Wang, Yiheng Zhu, Guoxian Song, Sizhe An, Janus Kristjansson, Xiao Yang, and Matthias Zwicker.
\newblock Panic-3d: Stylized single-view 3d reconstruction from portraits of anime characters.
\newblock In \emph{Proceedings of the IEEE/CVF Conference on Computer Vision and Pattern Recognition}, pages 21068--21077, 2023.

\bibitem[Ci et~al.(2018)Ci, Ma, Wang, Li, and Luo]{ci2018user}
Yuanzheng Ci, Xinzhu Ma, Zhihui Wang, Haojie Li, and Zhongxuan Luo.
\newblock User-guided deep anime line art colorization with conditional adversarial networks.
\newblock In \emph{Proceedings of the 26th ACM international conference on Multimedia}, pages 1536--1544, 2018.

\bibitem[Dai et~al.(2024)Dai, Zhou, Li, Li, and Loy]{dai2024learning}
Yuekun Dai, Shangchen Zhou, Qinyue Li, Chongyi Li, and Chen~Change Loy.
\newblock Learning inclusion matching for animation paint bucket colorization, 2024.

\bibitem[Deng et~al.(2009)Deng, Dong, Socher, Li, Li, and Fei-Fei]{deng2009imagenet}
Jia Deng, Wei Dong, Richard Socher, Li-Jia Li, Kai Li, and Li Fei-Fei.
\newblock Imagenet: A large-scale hierarchical image database.
\newblock In \emph{2009 IEEE conference on computer vision and pattern recognition}, pages 248--255. Ieee, 2009.

\bibitem[Dong et~al.(2015)Dong, Loy, He, and Tang]{dong2015image}
Chao Dong, Chen~Change Loy, Kaiming He, and Xiaoou Tang.
\newblock Image super-resolution using deep convolutional networks.
\newblock \emph{IEEE transactions on pattern analysis and machine intelligence}, 38\penalty0 (2):\penalty0 295--307, 2015.

\bibitem[Feng et~al.(2022)Feng, Zhai, Yang, Liang, Fan, Zhang, Shao, and Tao]{feng2022ic9600}
Tinglei Feng, Yingjie Zhai, Jufeng Yang, Jie Liang, Deng-Ping Fan, Jing Zhang, Ling Shao, and Dacheng Tao.
\newblock Ic9600: A benchmark dataset for automatic image complexity assessment.
\newblock \emph{IEEE Transactions on Pattern Analysis and Machine Intelligence}, 2022.

\bibitem[Goodfellow et~al.(2014)Goodfellow, Pouget-Abadie, Mirza, Xu, Warde-Farley, Ozair, Courville, and Bengio]{goodfellow2014generative}
Ian Goodfellow, Jean Pouget-Abadie, Mehdi Mirza, Bing Xu, David Warde-Farley, Sherjil Ozair, Aaron Courville, and Yoshua Bengio.
\newblock Generative adversarial nets.
\newblock \emph{Advances in neural information processing systems}, 27, 2014.

\bibitem[Guo et~al.(2023)Guo, Yang, Rao, Wang, Qiao, Lin, and Dai]{guo2023animatediff}
Yuwei Guo, Ceyuan Yang, Anyi Rao, Yaohui Wang, Yu Qiao, Dahua Lin, and Bo Dai.
\newblock Animatediff: Animate your personalized text-to-image diffusion models without specific tuning.
\newblock \emph{arXiv preprint arXiv:2307.04725}, 2023.

\bibitem[Han et~al.(2021)Han, Li, Mukherjee, Chiang, Grange, Chen, Su, Parker, Deng, Joshi, et~al.]{han2021technical}
Jingning Han, Bohan Li, Debargha Mukherjee, Ching-Han Chiang, Adrian Grange, Cheng Chen, Hui Su, Sarah Parker, Sai Deng, Urvang Joshi, et~al.
\newblock A technical overview of av1.
\newblock \emph{Proceedings of the IEEE}, 109\penalty0 (9):\penalty0 1435--1462, 2021.

\bibitem[Hati et~al.(2019)Hati, Jouet, Rousseaux, and Duhart]{hati2019paintstorch}
Yliess Hati, Gregor Jouet, Francis Rousseaux, and Cl{\'e}ment Duhart.
\newblock Paintstorch: a user-guided anime line art colorization tool with double generator conditional adversarial network.
\newblock In \emph{Proceedings of the 16th ACM SIGGRAPH European Conference on Visual Media Production}, pages 1--10, 2019.

\bibitem[He et~al.(2016)He, Zhang, Ren, and Sun]{he2016identity}
Kaiming He, Xiangyu Zhang, Shaoqing Ren, and Jian Sun.
\newblock Identity mappings in deep residual networks.
\newblock In \emph{Computer Vision--ECCV 2016: 14th European Conference, Amsterdam, The Netherlands, October 11--14, 2016, Proceedings, Part IV 14}, pages 630--645. Springer, 2016.

\bibitem[Huang et~al.(2023)Huang, Xie, Fukusato, and Miyata]{huang2023anifacedrawing}
Zhengyu Huang, Haoran Xie, Tsukasa Fukusato, and Kazunori Miyata.
\newblock Anifacedrawing: Anime portrait exploration during your sketching.
\newblock In \emph{ACM SIGGRAPH 2023 Conference Proceedings}, pages 1--11, 2023.

\bibitem[Isola et~al.(2017)Isola, Zhu, Zhou, and Efros]{isola2017image}
Phillip Isola, Jun-Yan Zhu, Tinghui Zhou, and Alexei~A Efros.
\newblock Image-to-image translation with conditional adversarial networks.
\newblock In \emph{Proceedings of the IEEE conference on computer vision and pattern recognition}, pages 1125--1134, 2017.

\bibitem[Ji et~al.(2020{\natexlab{a}})Ji, Cao, Tai, Wang, Li, and Huang]{Ji_2020_CVPR_Workshops}
Xiaozhong Ji, Yun Cao, Ying Tai, Chengjie Wang, Jilin Li, and Feiyue Huang.
\newblock Real-world super-resolution via kernel estimation and noise injection.
\newblock In \emph{The IEEE/CVF Conference on Computer Vision and Pattern Recognition (CVPR) Workshops}, 2020{\natexlab{a}}.

\bibitem[Ji et~al.(2020{\natexlab{b}})Ji, Cao, Tai, Wang, Li, and Huang]{ji2020real}
Xiaozhong Ji, Yun Cao, Ying Tai, Chengjie Wang, Jilin Li, and Feiyue Huang.
\newblock Real-world super-resolution via kernel estimation and noise injection.
\newblock In \emph{proceedings of the IEEE/CVF conference on computer vision and pattern recognition workshops}, pages 466--467, 2020{\natexlab{b}}.

\bibitem[Jiang et~al.(2023)Jiang, Jiang, Yang, and Loy]{jiang2023scenimefy}
Yuxin Jiang, Liming Jiang, Shuai Yang, and Chen~Change Loy.
\newblock Scenimefy: Learning to craft anime scene via semi-supervised image-to-image translation.
\newblock In \emph{Proceedings of the IEEE/CVF International Conference on Computer Vision}, pages 7357--7367, 2023.

\bibitem[Johnson et~al.(2016)Johnson, Alahi, and Fei-Fei]{johnson2016perceptual}
Justin Johnson, Alexandre Alahi, and Li Fei-Fei.
\newblock Perceptual losses for real-time style transfer and super-resolution.
\newblock In \emph{Computer Vision--ECCV 2016: 14th European Conference, Amsterdam, The Netherlands, October 11-14, 2016, Proceedings, Part II 14}, pages 694--711. Springer, 2016.

\bibitem[Kingma and Ba(2014)]{kingma2014adam}
Diederik~P Kingma and Jimmy Ba.
\newblock Adam: A method for stochastic optimization.
\newblock \emph{arXiv preprint arXiv:1412.6980}, 2014.

\bibitem[Lee and Lee(2020)]{lee2020automatic}
Yeongseop Lee and Seongjin Lee.
\newblock Automatic colorization of anime style illustrations using a two-stage generator.
\newblock \emph{Applied Sciences}, 10\penalty0 (23):\penalty0 8699, 2020.

\bibitem[Li et~al.(2023)Li, Fan, Xiang, Demandolx, Ranjan, Timofte, and Van~Gool]{li2023efficient}
Yawei Li, Yuchen Fan, Xiaoyu Xiang, Denis Demandolx, Rakesh Ranjan, Radu Timofte, and Luc Van~Gool.
\newblock Efficient and explicit modelling of image hierarchies for image restoration.
\newblock In \emph{Proceedings of the IEEE/CVF Conference on Computer Vision and Pattern Recognition}, pages 18278--18289, 2023.

\bibitem[Liang et~al.(2021)Liang, Cao, Sun, Zhang, Van~Gool, and Timofte]{liang2021swinir}
Jingyun Liang, Jiezhang Cao, Guolei Sun, Kai Zhang, Luc Van~Gool, and Radu Timofte.
\newblock Swinir: Image restoration using swin transformer.
\newblock In \emph{Proceedings of the IEEE/CVF international conference on computer vision}, pages 1833--1844, 2021.

\bibitem[Mitchell et~al.(1996)Mitchell, Pennebaker, Fogg, LeGall, Mitchell, Pennebaker, Fogg, and LeGall]{mitchell1996mpeg}
Joan~L Mitchell, William~B Pennebaker, Chad~E Fogg, Didier~J LeGall, Joan~L Mitchell, William~B Pennebaker, Chad~E Fogg, and Didier~J LeGall.
\newblock Mpeg-2 overview.
\newblock \emph{MPEG Video Compression Standard}, pages 171--186, 1996.

\bibitem[Mittal et~al.(2012{\natexlab{a}})Mittal, Moorthy, and Bovik]{mittal2012no}
Anish Mittal, Anush~Krishna Moorthy, and Alan~Conrad Bovik.
\newblock No-reference image quality assessment in the spatial domain.
\newblock \emph{IEEE Transactions on image processing}, 21\penalty0 (12):\penalty0 4695--4708, 2012{\natexlab{a}}.

\bibitem[Mittal et~al.(2012{\natexlab{b}})Mittal, Soundararajan, and Bovik]{mittal2012making}
Anish Mittal, Rajiv Soundararajan, and Alan~C Bovik.
\newblock Making a “completely blind” image quality analyzer.
\newblock \emph{IEEE Signal processing letters}, 20\penalty0 (3):\penalty0 209--212, 2012{\natexlab{b}}.

\bibitem[Miyato et~al.(2018)Miyato, Kataoka, Koyama, and Yoshida]{miyato2018spectral}
Takeru Miyato, Toshiki Kataoka, Masanori Koyama, and Yuichi Yoshida.
\newblock Spectral normalization for generative adversarial networks.
\newblock \emph{arXiv preprint arXiv:1802.05957}, 2018.

\bibitem[Schwarz et~al.(2007)Schwarz, Marpe, and Wiegand]{schwarz2007overview}
Heiko Schwarz, Detlev Marpe, and Thomas Wiegand.
\newblock Overview of the scalable video coding extension of the h. 264/avc standard.
\newblock \emph{IEEE Transactions on circuits and systems for video technology}, 17\penalty0 (9):\penalty0 1103--1120, 2007.

\bibitem[Shen et~al.(2022)Shen, Ming, Bao, Zhai, Chenn, and Gao]{shen2022enhanced}
Wang Shen, Cheng Ming, Wenbo Bao, Guangtao Zhai, Li Chenn, and Zhiyong Gao.
\newblock Enhanced deep animation video interpolation.
\newblock In \emph{2022 IEEE International Conference on Image Processing (ICIP)}, pages 31--35. IEEE, 2022.

\bibitem[Si and Shen(2016)]{si2016research}
Zhanjun Si and Ke Shen.
\newblock Research on the webp image format.
\newblock In \emph{Advanced graphic communications, packaging technology and materials}, pages 271--277. Springer, 2016.

\bibitem[Simonyan and Zisserman(2014)]{simonyan2014very}
Karen Simonyan and Andrew Zisserman.
\newblock Very deep convolutional networks for large-scale image recognition.
\newblock \emph{arXiv preprint arXiv:1409.1556}, 2014.

\bibitem[Siyao et~al.(2021)Siyao, Zhao, Yu, Sun, Metaxas, Loy, and Liu]{siyao2021deep}
Li Siyao, Shiyu Zhao, Weijiang Yu, Wenxiu Sun, Dimitris Metaxas, Chen~Change Loy, and Ziwei Liu.
\newblock Deep animation video interpolation in the wild.
\newblock In \emph{Proceedings of the IEEE/CVF conference on computer vision and pattern recognition}, pages 6587--6595, 2021.

\bibitem[Siyao et~al.(2022)Siyao, Li, Li, Dong, Liu, and Loy]{siyao2022animerun}
Li Siyao, Yuhang Li, Bo Li, Chao Dong, Ziwei Liu, and Chen~Change Loy.
\newblock Animerun: 2d animation visual correspondence from open source 3d movies.
\newblock \emph{Advances in Neural Information Processing Systems}, 35:\penalty0 18996--19007, 2022.

\bibitem[Siyao et~al.(2023)Siyao, Gu, Xiao, Ding, Liu, and Loy]{siyao2023deep}
Li Siyao, Tianpei Gu, Weiye Xiao, Henghui Ding, Ziwei Liu, and Chen~Change Loy.
\newblock Deep geometrized cartoon line inbetweening.
\newblock In \emph{Proceedings of the IEEE/CVF International Conference on Computer Vision}, pages 7291--7300, 2023.

\bibitem[Su et~al.(2020)Su, Yan, Zhu, Zhang, Ge, Sun, and Zhang]{su2020blindly}
Shaolin Su, Qingsen Yan, Yu Zhu, Cheng Zhang, Xin Ge, Jinqiu Sun, and Yanning Zhang.
\newblock Blindly assess image quality in the wild guided by a self-adaptive hyper network.
\newblock In \emph{Proceedings of the IEEE/CVF Conference on Computer Vision and Pattern Recognition}, pages 3667--3676, 2020.

\bibitem[Sullivan et~al.(2012)Sullivan, Ohm, Han, and Wiegand]{sullivan2012overview}
Gary~J Sullivan, Jens-Rainer Ohm, Woo-Jin Han, and Thomas Wiegand.
\newblock Overview of the high efficiency video coding (hevc) standard.
\newblock \emph{IEEE Transactions on circuits and systems for video technology}, 22\penalty0 (12):\penalty0 1649--1668, 2012.

\bibitem[Timofte et~al.(2017)Timofte, Agustsson, Van~Gool, Yang, and Zhang]{timofte2017ntire}
Radu Timofte, Eirikur Agustsson, Luc Van~Gool, Ming-Hsuan Yang, and Lei Zhang.
\newblock Ntire 2017 challenge on single image super-resolution: Methods and results.
\newblock In \emph{Proceedings of the IEEE conference on computer vision and pattern recognition workshops}, pages 114--125, 2017.

\bibitem[Tuo et~al.(2023)Tuo, Yang, Fu, Dun, and Qian]{tuo2023learning}
Zixi Tuo, Huan Yang, Jianlong Fu, Yujie Dun, and Xueming Qian.
\newblock Learning data-driven vector-quantized degradation model for animation video super-resolution.
\newblock \emph{arXiv preprint arXiv:2303.09826}, 2023.

\bibitem[Wallace(1992)]{wallace1992jpeg}
Gregory~K Wallace.
\newblock The jpeg still picture compression standard.
\newblock \emph{IEEE transactions on consumer electronics}, 38\penalty0 (1):\penalty0 xviii--xxxiv, 1992.

\bibitem[Wang et~al.(2024)Wang, Liu, Liu, and Yang]{wang2024vcisr}
Boyang Wang, Bowen Liu, Shiyu Liu, and Fengyu Yang.
\newblock Vcisr: Blind single image super-resolution with video compression synthetic data.
\newblock In \emph{Proceedings of the IEEE/CVF Winter Conference on Applications of Computer Vision}, pages 4302--4312, 2024.

\bibitem[Wang et~al.(2023{\natexlab{a}})Wang, Chan, and Loy]{wang2023exploring}
Jianyi Wang, Kelvin~CK Chan, and Chen~Change Loy.
\newblock Exploring clip for assessing the look and feel of images.
\newblock In \emph{Proceedings of the AAAI Conference on Artificial Intelligence}, pages 2555--2563, 2023{\natexlab{a}}.

\bibitem[Wang et~al.(2023{\natexlab{b}})Wang, Niu, Dou, Wang, Wang, Ming, Liu, and Li]{wang2023coloring}
Ning Wang, Muyao Niu, Zhi Dou, Zhihui Wang, Zhiyong Wang, Zhaoyan Ming, Bin Liu, and Haojie Li.
\newblock Coloring anime line art videos with transformation region enhancement network.
\newblock \emph{Pattern Recognition}, 141:\penalty0 109562, 2023{\natexlab{b}}.

\bibitem[Wang et~al.(2018{\natexlab{a}})Wang, Liu, Zhu, Tao, Kautz, and Catanzaro]{wang2018high}
Ting-Chun Wang, Ming-Yu Liu, Jun-Yan Zhu, Andrew Tao, Jan Kautz, and Bryan Catanzaro.
\newblock High-resolution image synthesis and semantic manipulation with conditional gans.
\newblock In \emph{Proceedings of the IEEE conference on computer vision and pattern recognition}, pages 8798--8807, 2018{\natexlab{a}}.

\bibitem[Wang et~al.(2018{\natexlab{b}})Wang, Yu, Dong, and Loy]{wang2018recovering}
Xintao Wang, Ke Yu, Chao Dong, and Chen~Change Loy.
\newblock Recovering realistic texture in image super-resolution by deep spatial feature transform.
\newblock In \emph{Proceedings of the IEEE conference on computer vision and pattern recognition}, pages 606--615, 2018{\natexlab{b}}.

\bibitem[Wang et~al.(2018{\natexlab{c}})Wang, Yu, Wu, Gu, Liu, Dong, Qiao, and Change~Loy]{wang2018esrgan}
Xintao Wang, Ke Yu, Shixiang Wu, Jinjin Gu, Yihao Liu, Chao Dong, Yu Qiao, and Chen Change~Loy.
\newblock Esrgan: Enhanced super-resolution generative adversarial networks.
\newblock In \emph{Proceedings of the European conference on computer vision (ECCV) workshops}, pages 0--0, 2018{\natexlab{c}}.

\bibitem[Wang et~al.(2021)Wang, Xie, Dong, and Shan]{wang2021realesrgan}
Xintao Wang, Liangbin Xie, Chao Dong, and Ying Shan.
\newblock Real-esrgan: Training real-world blind super-resolution with pure synthetic data.
\newblock In \emph{Proceedings of the IEEE/CVF international conference on computer vision}, pages 1905--1914, 2021.

\bibitem[Winnem{\"o}ller et~al.(2012)Winnem{\"o}ller, Kyprianidis, and Olsen]{winnemoller2012xdog}
Holger Winnem{\"o}ller, Jan~Eric Kyprianidis, and Sven~C Olsen.
\newblock Xdog: An extended difference-of-gaussians compendium including advanced image stylization.
\newblock \emph{Computers \& Graphics}, 36\penalty0 (6):\penalty0 740--753, 2012.

\bibitem[Wu et~al.(2022)Wu, Wang, Li, and Shan]{wu2022animesr}
Yanze Wu, Xintao Wang, Gen Li, and Ying Shan.
\newblock Animesr: Learning real-world super-resolution models for animation videos.
\newblock \emph{arXiv preprint arXiv:2206.07038}, 2022.

\bibitem[Xiao et~al.(2020)Xiao, Xiong, Fu, Liu, and Zha]{xiao2020space}
Zeyu Xiao, Zhiwei Xiong, Xueyang Fu, Dong Liu, and Zheng-Jun Zha.
\newblock Space-time video super-resolution using temporal profiles.
\newblock In \emph{Proceedings of the 28th ACM International Conference on Multimedia}, pages 664--672, 2020.

\bibitem[Xiao et~al.(2021)Xiao, Fu, Huang, Cheng, and Xiong]{xiao2021space}
Zeyu Xiao, Xueyang Fu, Jie Huang, Zhen Cheng, and Zhiwei Xiong.
\newblock Space-time distillation for video super-resolution.
\newblock In \emph{Proceedings of the IEEE/CVF conference on computer vision and pattern recognition}, pages 2113--2122, 2021.

\bibitem[Xiao et~al.(2023{\natexlab{a}})Xiao, Bai, Lu, and Xiong]{xiao2023dive}
Zeyu Xiao, Jiawang Bai, Zhihe Lu, and Zhiwei Xiong.
\newblock A dive into sam prior in image restoration.
\newblock \emph{arXiv preprint arXiv:2305.13620}, 2023{\natexlab{a}}.

\bibitem[Xiao et~al.(2023{\natexlab{b}})Xiao, Liu, Gao, and Xiong]{xiao2023cutmib}
Zeyu Xiao, Yutong Liu, Ruisheng Gao, and Zhiwei Xiong.
\newblock Cutmib: Boosting light field super-resolution via multi-view image blending.
\newblock In \emph{Proceedings of the IEEE/CVF Conference on Computer Vision and Pattern Recognition}, pages 1672--1682, 2023{\natexlab{b}}.

\bibitem[Xu et~al.(2022)Xu, Dutta, He, and Matsumaru]{xu2022transformer}
Shizhuo Xu, Vibekananda Dutta, Xin He, and Takafumi Matsumaru.
\newblock A transformer-based model for super-resolution of anime image.
\newblock \emph{Sensors}, 22\penalty0 (21):\penalty0 8126, 2022.

\bibitem[Yang et~al.(2022)Yang, Wu, Shi, Lao, Gong, Cao, Wang, and Yang]{yang2022maniqa}
Sidi Yang, Tianhe Wu, Shuwei Shi, Shanshan Lao, Yuan Gong, Mingdeng Cao, Jiahao Wang, and Yujiu Yang.
\newblock Maniqa: Multi-dimension attention network for no-reference image quality assessment.
\newblock In \emph{Proceedings of the IEEE/CVF Conference on Computer Vision and Pattern Recognition}, pages 1191--1200, 2022.

\bibitem[Yao et~al.(2016)Yao, Hung, Li, Chen, Adhitya, and Lai]{yao2016manga}
Chih-Yuan Yao, Shih-Hsuan Hung, Guo-Wei Li, I-Yu Chen, Reza Adhitya, and Yu-Chi Lai.
\newblock Manga vectorization and manipulation with procedural simple screentone.
\newblock \emph{IEEE transactions on visualization and computer graphics}, 23\penalty0 (2):\penalty0 1070--1084, 2016.

\bibitem[Yuan et~al.(2018)Yuan, Liu, Zhang, Zhang, Dong, and Lin]{yuan2018unsupervised}
Yuan Yuan, Siyuan Liu, Jiawei Zhang, Yongbing Zhang, Chao Dong, and Liang Lin.
\newblock Unsupervised image super-resolution using cycle-in-cycle generative adversarial networks.
\newblock In \emph{Proceedings of the IEEE conference on computer vision and pattern recognition workshops}, pages 701--710, 2018.

\bibitem[Zhang et~al.(2021{\natexlab{a}})Zhang, Liang, Van~Gool, and Timofte]{zhang2021designing}
Kai Zhang, Jingyun Liang, Luc Van~Gool, and Radu Timofte.
\newblock Designing a practical degradation model for deep blind image super-resolution.
\newblock In \emph{Proceedings of the IEEE/CVF International Conference on Computer Vision}, pages 4791--4800, 2021{\natexlab{a}}.

\bibitem[Zhang et~al.(2021{\natexlab{b}})Zhang, Li, Simo-Serra, Ji, Wong, and Liu]{zhang2021user}
Lvmin Zhang, Chengze Li, Edgar Simo-Serra, Yi Ji, Tien-Tsin Wong, and Chunping Liu.
\newblock User-guided line art flat filling with split filling mechanism.
\newblock In \emph{Proceedings of the IEEE/CVF conference on computer vision and pattern recognition}, pages 9889--9898, 2021{\natexlab{b}}.

\bibitem[Zhang et~al.(2023)Zhang, Rao, and Agrawala]{zhang2023adding}
Lvmin Zhang, Anyi Rao, and Maneesh Agrawala.
\newblock Adding conditional control to text-to-image diffusion models.
\newblock In \emph{Proceedings of the IEEE/CVF International Conference on Computer Vision}, pages 3836--3847, 2023.

\bibitem[Zhang et~al.(2009)Zhang, Chen, Zhang, Hu, and Martin]{zhang2009vectorizing}
Song-Hai Zhang, Tao Chen, Yi-Fei Zhang, Shi-Min Hu, and Ralph~R Martin.
\newblock Vectorizing cartoon animations.
\newblock \emph{IEEE Transactions on Visualization and Computer Graphics}, 15\penalty0 (4):\penalty0 618--629, 2009.

\bibitem[Zhao et~al.(2022)Zhao, Ren, Chen, Jia, Wang, and Liu]{zhao2022cartoon}
Yang Zhao, Diya Ren, Yuan Chen, Wei Jia, Ronggang Wang, and Xiaoping Liu.
\newblock Cartoon image processing: A survey.
\newblock \emph{IJCV}, 2022.

\end{thebibliography}
}

\clearpage
\setcounter{page}{1}
\maketitlesupplementary
In this supplementary material, Sec.~\ref{sec:API_details} first presents more statistics and details of our proposed anime image SR training dataset. 
Then, Sec.~\ref{sec:implementation_details} shows details about our implementations in super-resolution (SR) network training. Specifically, Sec.~\ref{subsec:network} presents the image SR network we used in our training. Sec.~\ref{subsec:hand_drawn_line} presents details of post-processing techniques we use on the pseudo-GT preparation for hand-drawn line enhancement. Sec.~\ref{subsec:perceptual_loss} presents figures and details of the ResNet50~\cite{he2016identity} perceptual loss for our proposed balanced twin perceptual loss. Sec.~\ref{subsec:degradation_details} provides the hyperparameter setting for our proposed prediction-oriented compression and shuffled resize module in the degradation model.
Finally, Sec.~\ref{sec:more_qualitative} provides more visual results of comparisons among SOTA methods and ablation studies.

\appendix
\section{API Dataset Details}
\label{sec:API_details}
\begin{figure}[t]
  \centering
  \includegraphics[width= 1.0\columnwidth]{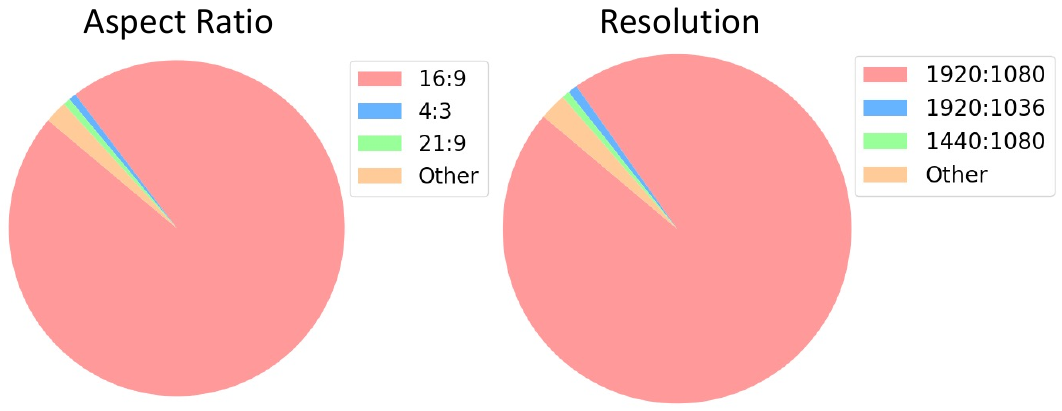}

   \caption{
     \textbf{ API dataset extra statistics.}
   }

   \vspace{-0.2cm}
   
    \label{fig:API_Statistics}
\end{figure}

Our \textbf{A}nime \textbf{P}roduction-oriented \textbf{I}mage (API) SR dataset contains 3,740 high-quality and informative images. This quantity is roughly the same quantity as the previous photorealistic SR training dataset size~\cite{wang2021realesrgan, zhang2021designing}, which includes DIV2K~\cite{agustsson2017ntire}, Flickr2K~\cite{timofte2017ntire}, and OutdoorSceneTraining~\cite{wang2018recovering}.
The aspect ratio and resolution information before scaling are shown in Fig.~\ref{fig:API_Statistics}.

\section{Implementation Details}
\label{sec:implementation_details}

\subsection{Training Network Details}
\label{subsec:network}
The generator network we deploy is GRL~\cite{li2023efficient}, a SOTA image SR network (CVPR 2023). 
GRL leverages interconnected relationships within various layers of image structures through a Transformer-based framework, attaining improvement in multiple tasks of SR and image restoration.
The model we chose is its tiny version, which has 0.91M parameters. 
To better adapt the real-world SR task, we changed its upsampler module from the default pixel shuffle strategy to the nearest neighbor interpolation with the convolution layer approach, which is used for the base model version but not for the tiny version in their proposed methods. 
We change the upsampler because the nearest neighbor interpolation with the convolution layer is claimed to show fewer artifacts in the upsampling process than the pixel shuffle strategy. 
The final network parameter is 1.03M, which is the smallest network among all image and video-based SOTA methods that we compare.

\subsection{Hand-drawn line enhancement Details}
\label{subsec:hand_drawn_line}
In the hand-drawn line enhancement, we have proposed outlier filter and passive dilate techniques to obtain a clean XDoG-extracted~\cite{winnemoller2012xdog} hand-drawn line edge map.
XDoG is widely used in paired dataset preparation in anime colorization~\cite{carrillo2023diffusart, cao2023animediffusion, wang2023coloring, huang2023anifacedrawing}.
The extracted edge map by XDoG is a binary output, where the white pixel stands for the active edge map region and the black pixel stands for the unrelated region.

For the outlier filter, we use breadth-first search in eight directions to recursively detect the surrounding pixels of all white pixels and turn white pixel regions into black pixels if the total quantity of connected white pixels is less than the threshold. We empirically set the threshold as 32.

\begin{figure}[t]
  \centering
  \includegraphics[width= 1.0\columnwidth]{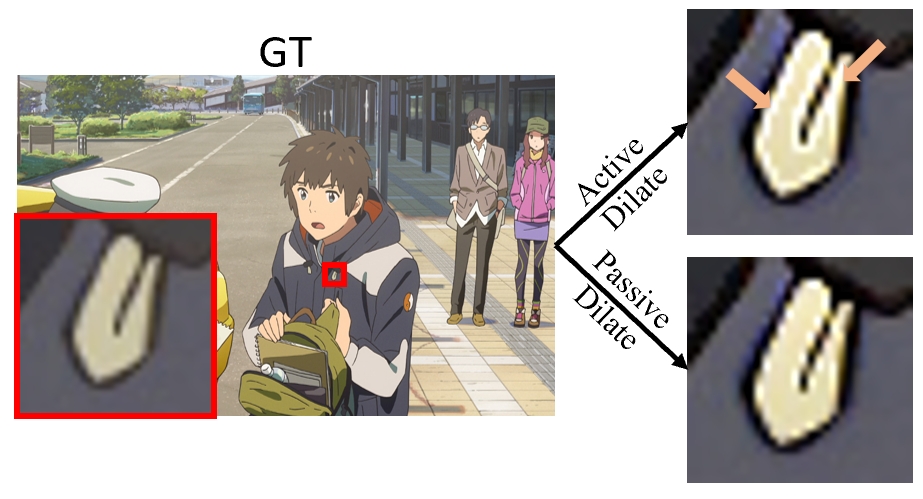}

  
   \caption{
     \textbf{Comparisons between active and passive dilation.} Our proposed passive dilation is more concentrated on the hand-drawn line region without producing over-sharpened pseudo-GT images as in active dilation methods.
   }

   \vspace{-0.2cm}
   
    \label{fig:dilate}
\end{figure}

For the dilation, we passively replace the black pixel with the white pixel if it has more than 3 white pixel neighbors, which is different from independent kernel-based active dilation methods in~\cite{lee2020automatic, ci2018user, hati2019paintstorch} that directly spread the surrounding neighbors to be white pixels if the central pixel is white. 
Compared to active dilation methods, our proposed passive dilation is more concentrated on the hand-drawn lines region instead of covering unrelated pixel information (see Fig.~\ref{fig:dilate}). Thus, we name our methods as passive dilatation.

In the implementation, we will do an unsharp mask for the whole image first to increase overall visualization sharpness and then apply two extra turns of sharpening to the hand-drawn lines specifically based on the pipeline design mentioned above. 
More implementation details can be found in our released code.

\subsection{Balanced Twin Perceptual Loss Details}
\label{subsec:perceptual_loss}
\begin{figure*}[t]
  \centering
  
  
  \includegraphics[width=\linewidth]{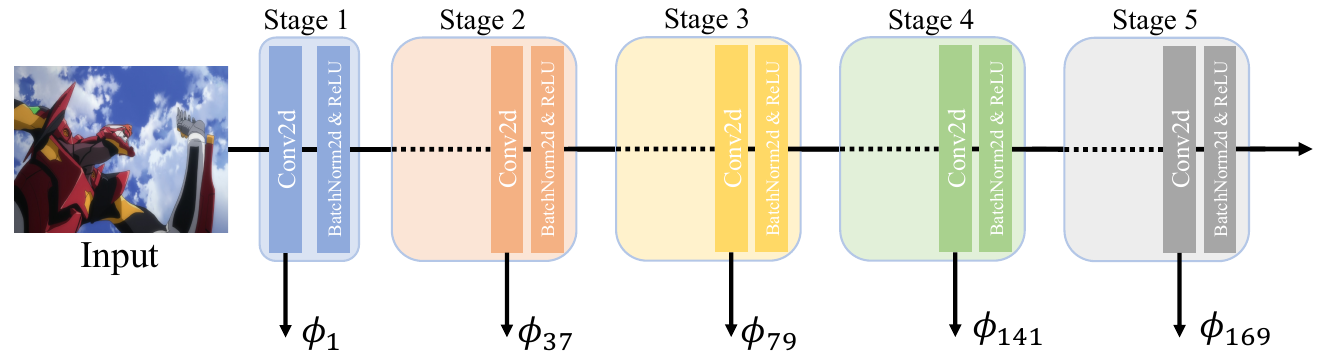}
  
  
   \caption{
   The overview of our proposed middle-layer outputs of ResNet50~\cite{he2016identity} perceptual loss trained by Danbooru dataset~\cite{danbooru2018resnet}.
   Overall, ResNet50 can be summarized into five stages which is similar to VGG~\cite{simonyan2014very}.
   $\phi_j$ represents the perceptual function that returns $j$th layer output of ResNet50.
   }
   
   \vspace{-0.1cm}
   \label{fig:ResNet_Perceptual}
\end{figure*}

As shown in Fig.~\ref{fig:ResNet_Perceptual}, our proposed middle-layer output comparisons for ResNet50~\cite{he2016identity} follow the idea proposed by ESRGAN~\cite{wang2018esrgan} which compares feature map outputs before the activation layer. Following VGG-based perceptual loss~\cite{johnson2016perceptual}, we compare the last convolution layer of each stage.
There are five middle-layer output comparisons, which are the same quantity as VGG-based perceptual loss~\cite{johnson2016perceptual}. Thus, our proposed twin perpetual loss reaches a mutual balance in training.


\subsection{Degradation Details}
\label{subsec:degradation_details}
For the prediction-oriented compression module of the degradation model, we deploy both the image compression with prediction mechanism (\ie, WebP~\cite{si2016research} and AVIF~\cite{han2021technical}) and single-frame video compression. 
Meanwhile, for the robustness of the degradation model, we keep the JPEG~\cite{wallace1992jpeg}. The quality factor range of JPEG, WebP, and AVIF is $[20, 95]$ with encoding speed in the range of $[0, 6]$ for WebP and AVIF. The probability of fetching the value in the range is equal.

For the stability of video compression processing, we choose the widely-used video processing tools, \textit{ffmpeg}, to perform the proposed single-frame compression of MPEG2~\cite{mitchell1996mpeg}, MPEG4~\cite{avaro2000mpeg}, H.264~\cite{schwarz2007overview}, and H.265~\cite{sullivan2012overview}. 
In \textit{ffmpeg}, CRF is an engineering system to control the quantization level, and preset is a speed control mechanism whose setting is directly related to compression distortions. For MPEG2 and MPEG4, we empirically find that the quality factor control  (\textit{-qscale:v}) is a better way to control single-frame compression, but for H.264 and H.265, CRF is a better way to control.
For MPEG2 and MPEG4, we set the quality factor in the range $[8, 31]$. For H.264 and H.265, we set the CRF in the range $[23, 38]$ and $[28, 42]$ respectively. 
The preset for all of them is $\{slow, medium, fast, faster, superfast\}$ with probability $\{0.05, 0.35, 0.3, 0.2, 0.1\}$.

The first prediction-oriented compression includes JPEG~\cite{wallace1992jpeg} and WebP~\cite{si2016research} with a probability of $\{0.4, 0.6\}$ respectively. The second prediction-oriented compression includes JPEG~\cite{wallace1992jpeg}, WebP~\cite{si2016research}, AVIF~\cite{han2021technical}, and single-frame compression of MPEG2~\cite{mitchell1996mpeg}, MPEG4~\cite{avaro2000mpeg}, H.264~\cite{schwarz2007overview}, and H.265~\cite{sullivan2012overview} with probability of $\{0.06, 0.1, 0.1, 0.12, 0.12, 0.3, 0.2\}$ respectively.  
For the first resize module, we set the scaling in the range of $[0.1, 1.2]$ with probability $\{0.2, 0.7, 0.1\}$ to scale up, scale down, or remain current resolution. For the second resize module, we choose the range of $[0.15, 1.2]$ with probability $\{0.2, 0.7, 0.1\}$.
More implementation details can be found in our released code.

\section{More Qualitative Comparisons}
\label{sec:more_qualitative}
In this section, we present more qualitative results to verify the effectiveness of our APISR among SOTA methods. Moreover, we provide visual comparisons for the ablation studies.

\begin{figure*}[t]
  \centering
  
  \includegraphics[width=\linewidth]{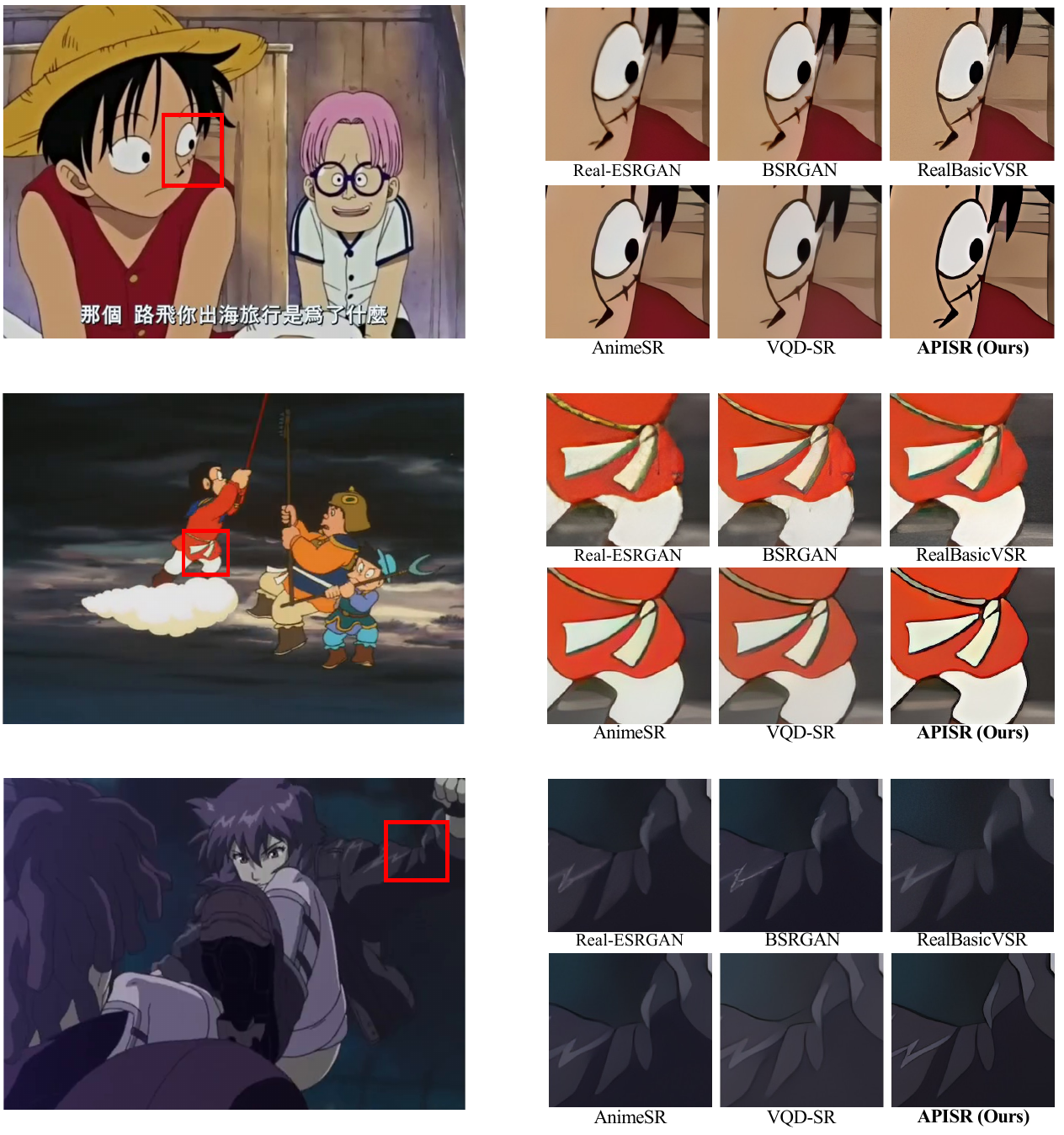}

  
   \caption{
   Qualitative comparisons on AVC-RealLQ~\cite{wu2022animesr} for $4\times$ scaling. 
   Our APISR presents clearer and sharper hand-drawn lines, better restoration with more natural details, and does not present unwanted color artifacts.
   \textbf{Zoom in for the best view.}
   }

   \label{fig:Supp_visual1}
\end{figure*}

\begin{figure*}[t]
  \centering
  
  \includegraphics[width=\linewidth]{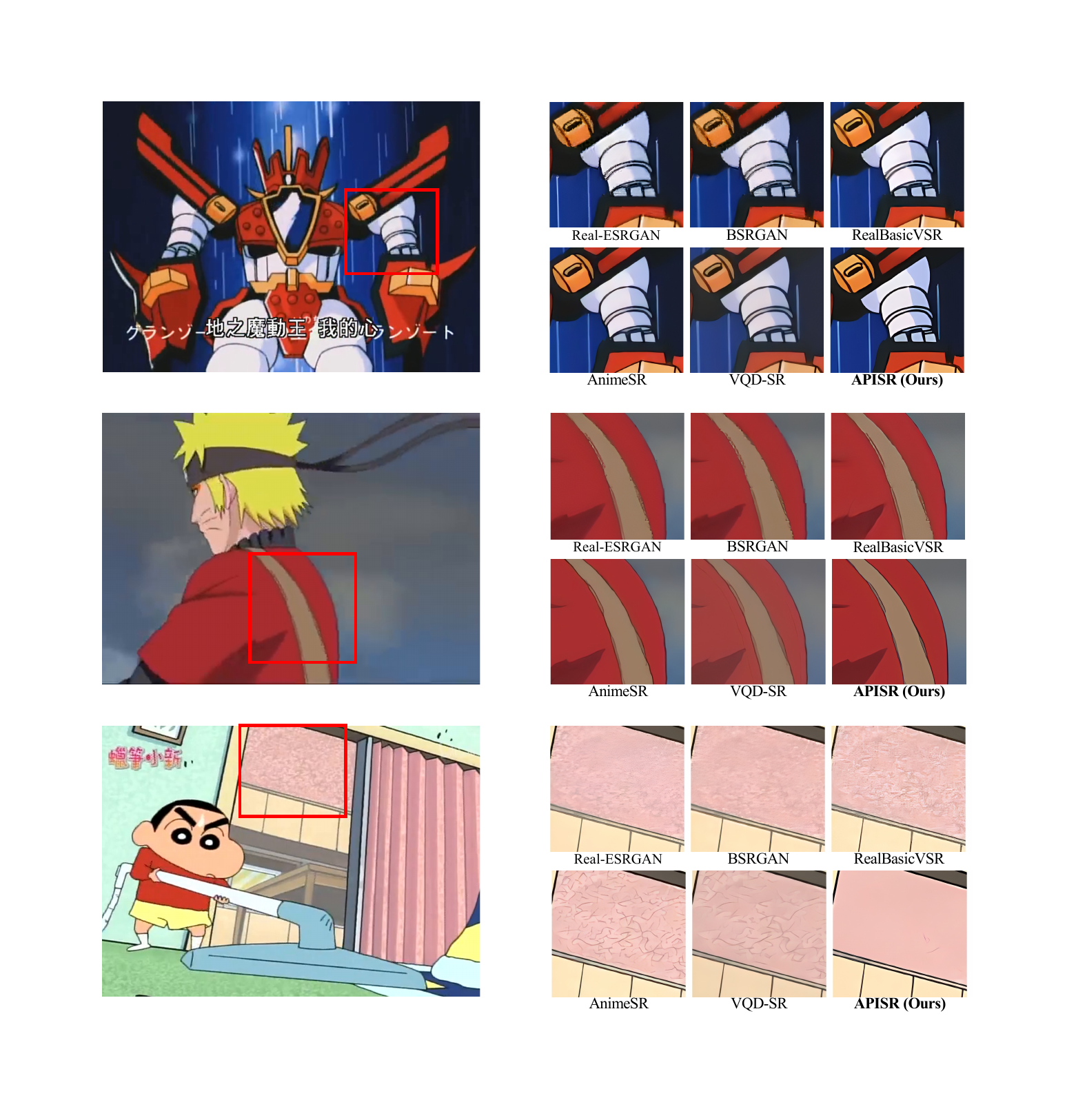}

  
   \caption{
   Qualitative comparisons on AVC-RealLQ~\cite{wu2022animesr} for $4\times$ scaling. 
   Our APISR presents clearer and sharper hand-drawn lines, better restoration with more natural details, and does not present unwanted color artifacts.
   \textbf{Zoom in for the best view.}
   }

   \label{fig:Supp_visual2}
\end{figure*}

\begin{figure*}[t]
  \centering
  
  \includegraphics[width=\linewidth]{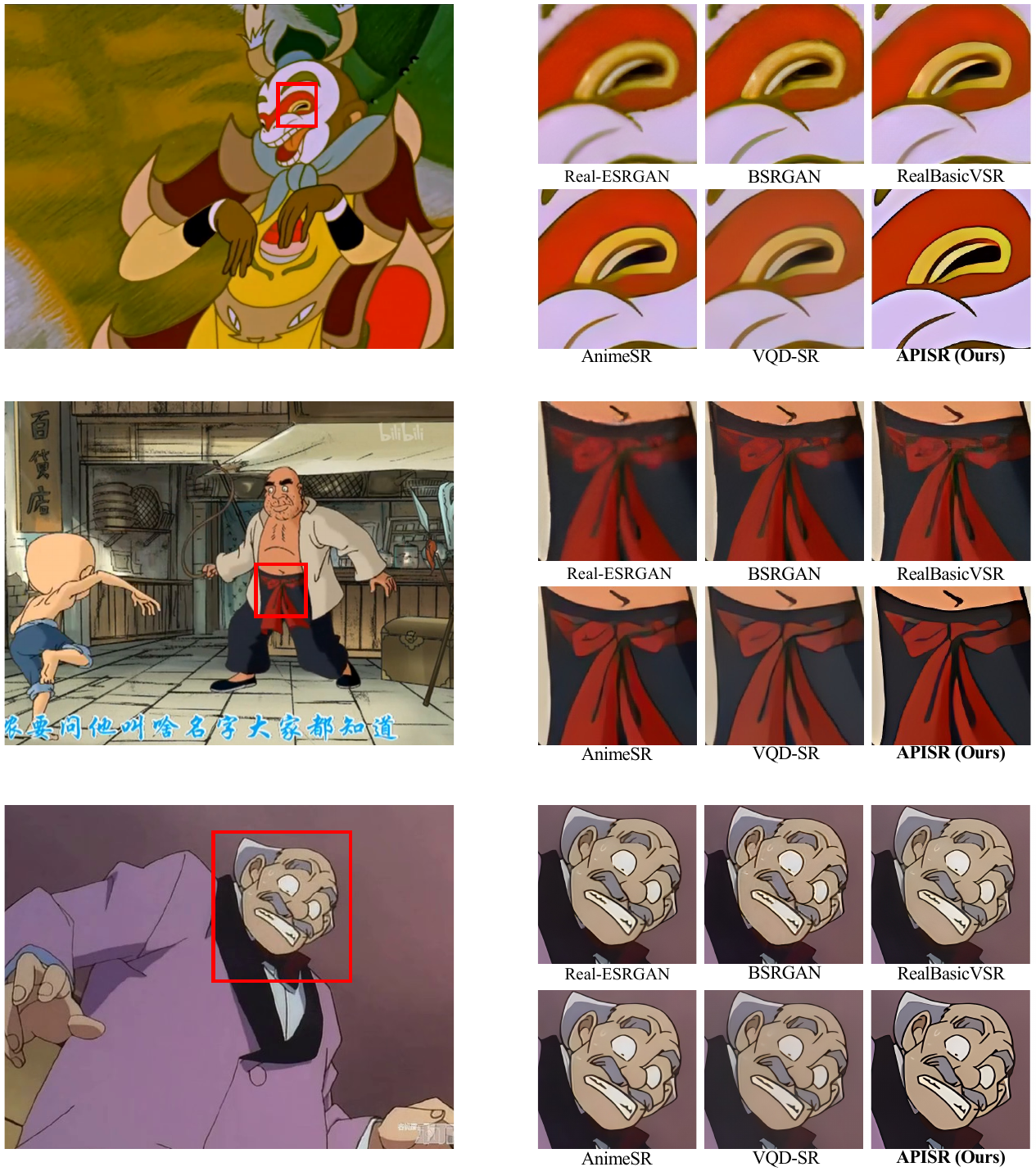}

  
   \caption{
   Qualitative comparisons on AVC-RealLQ~\cite{wu2022animesr} for $4\times$ scaling. 
   Our APISR presents clearer and sharper hand-drawn lines, better restoration with more natural details, and does not present unwanted color artifacts.
   \textbf{Zoom in for the best view.}
   }

   \label{fig:Supp_visual3}
\end{figure*}

\mypar{Extra Qualitative Comparisons with SOTA methods.}
Fig.~\ref{fig:Supp_visual1} and Fig.~\ref{fig:Supp_visual2} show extra qualitative comparisons on AVC-RealLQ~\cite{wu2022animesr} datasets for $4\times$ scaling. 
This includes image-based Real-ESRGAN~\cite{wang2021realesrgan} and BSRGAN~\cite{zhang2021designing}, and video-based RealBasicVSR~\cite{chan2022investigating}, AnimeSR~\cite{wu2022animesr}, and VQD-SR~\cite{tuo2023learning}.
Our APISR presents clearer and sharper hand-drawn lines (first example of Fig.~\ref{fig:Supp_visual1}, first and second examples of Fig.~\ref{fig:Supp_visual2}, and third example of Fig.~\ref{fig:Supp_visual3}),
better restoration with more natural details (second and third examples of Fig.~\ref{fig:Supp_visual1}, and third example of Fig.~\ref{fig:Supp_visual2}), 
and does not present unwanted color artifacts (first and second examples of Fig.~\ref{fig:Supp_visual3}).

\begin{figure*}[t]
     \vspace{0cm}
  \centering
  \includegraphics[width=\linewidth]{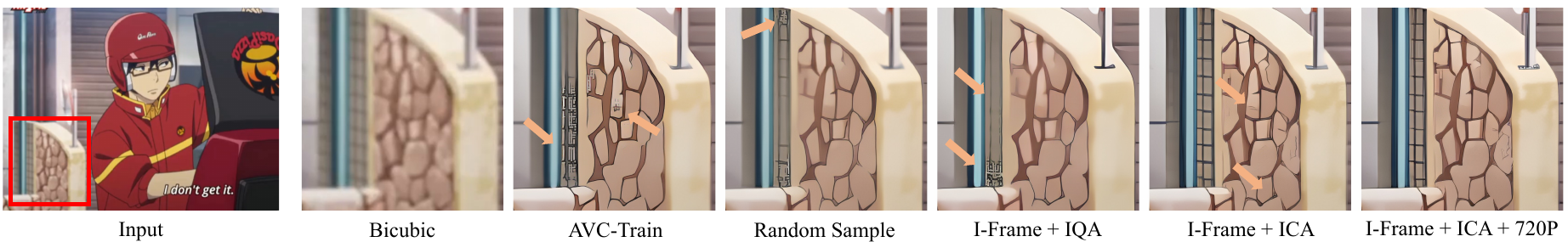}

   \caption{
   \textbf{Qualitative comparisons of the first ablation study.}
   IQA stands for image quality assessment. ICA stands for image complexity assessment. 720P stands for our proposed 720P rescaling.
   \textbf{Zoom in for the best view.}
   }

   \label{fig:ablation1}

\end{figure*}

\begin{figure*}[t]

     \vspace{-0cm}
  \centering
  \includegraphics[width=\linewidth]{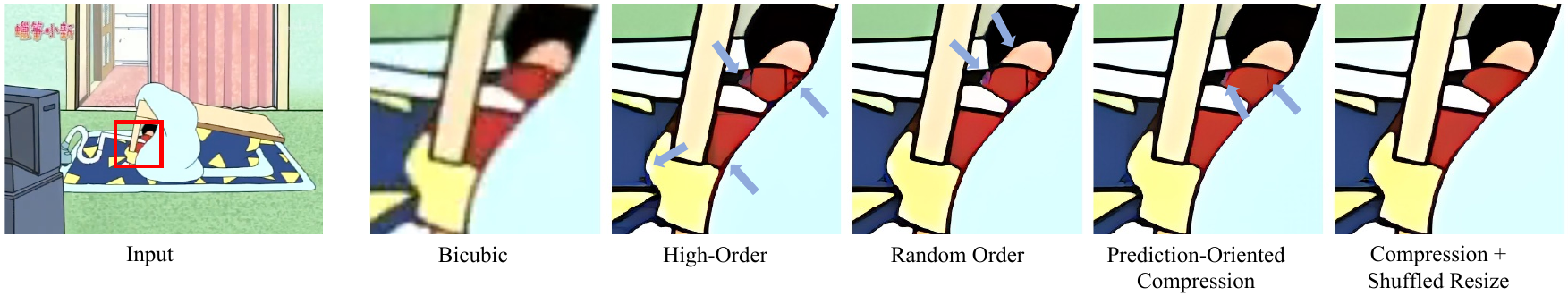}

   \caption{
        \textbf{Qualitative comparisons of the second ablation study.}
        \textbf{Zoom in for the best view.}
   }

   \label{fig:ablation2}

\end{figure*}

\begin{figure*}[t]

     \vspace{-0.0cm}
  \centering
  \includegraphics[width=\linewidth]{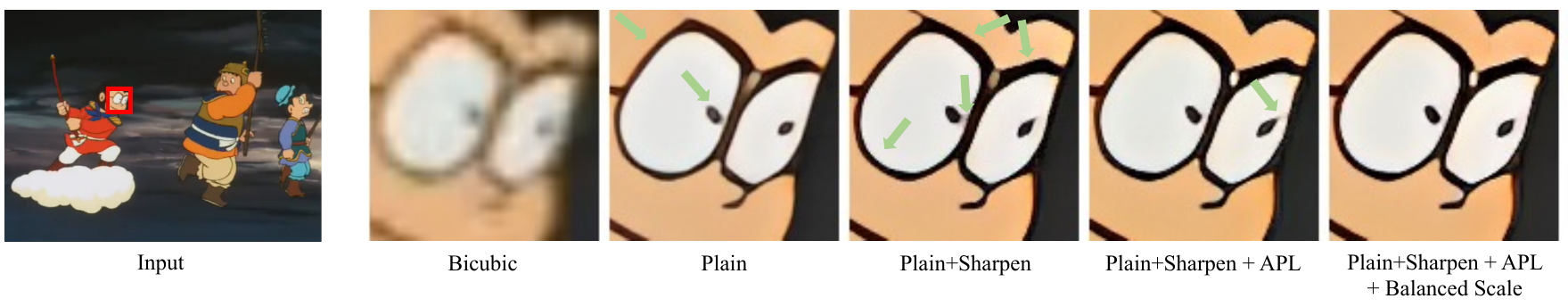}
  
   
   \caption{
   \textbf{Qualitative comparisons of the third ablation study.}
   Hand-drawn lines enhancement is denoted as \textbf{Sharpen} and twin perceptual loss is denoted as \textbf{APL}. \textbf{Balanced Scale} presents the early layer scaling to ResNet perceptual loss.
   \textbf{Zoom in for the best view.}
   }

   \label{fig:ablation3}

\end{figure*}

\mypar{Qualitative Comparisons of Ablation Studies.}
Fig.~\ref{fig:ablation1}, Fig.~\ref{fig:ablation2}, and Fig.~\ref{fig:ablation3} shows the qualitative comparisons of ablations studies.

As shown in Fig.~\ref{fig:ablation1}, the network trained with AVC-Train~\cite{wu2022animesr} over-sharpens the grid texture and produces annoying artifacts as denoted by the arrows in the figure. 
Similarly, the network trained with the random sampled or IQA-based sampled dataset can alleviate this artifact but is still hard to completely remove it. However, when we introduce the ICA-based selection method with I-Frame dataset collection, this artifact is greatly removed and the generated image shows more natural details. This is thanks to versatile complex scenes included in the dataset due to ICA-based selection. With 720P rescaling, fewer ringing artifacts appear.

As shown in Fig.~\ref{fig:ablation2}, the network trained with high-order~\cite{wang2021realesrgan} and random order~\cite{zhang2021designing} degradation model presents ringing artifacts, rainbow effects, and color distortions as denoted by the arrows in the figure. 
Nevertheless, introducing our proposed prediction-oriented compression module in the degradation model promotes the network to greatly restore these problems and generate more natural details with less distorted hand-drawn lines. 
Moreover, with the shuffled resize module in the degradation model, more distortions are restored and present natural shadow details.

As shown in Fig.~\ref{fig:ablation3}, the network trained with the plain version presents unwanted color pixel artifacts and sparse hand-drawn line information as denoted by the arrows in the figure.
With the hand-drawn line enhancement, the hand-drawn line around the eyes of the character is greatly intensified and more details are generated. However, the unwanted color pixels still exist and they are presented as an annoying artifact. With the twin perceptual loss, the unwanted color pixels are greatly alleviated. Further, with the scaling to early layers in ResNet perceptual loss, more shadow artifacts and distortions are restored.

\clearpage

\clearpage

\end{document}